\documentclass[%
 reprint,
 amsmath,amssymb,
 aps,
]{revtex4-2}

\usepackage{graphicx}
\usepackage{dcolumn}
\usepackage{bm}
\usepackage{color}
\usepackage{soul}
\usepackage{placeins}
\usepackage{makecell}
\usepackage{booktabs}
\usepackage{longtable}
\usepackage{rotating}
\usepackage{cancel}
\usepackage{url}

\begin{document}

\preprint{APS/123-QED}
\title{
Using firm-level supply chain networks to measure the speed of the energy transition
}

\author{Johannes Stangl}
\affiliation{Complexity Science Hub, Metternichgasse 8, 1030 Wien}

\author{András Borsos}%
\affiliation{Complexity Science Hub, Metternichgasse 8, 1030 Wien}
\affiliation{National Bank of Hungary, Szabadság tér 9, 1054 Budapest}
\affiliation{Institute for New Economic Thinking, Manor Road Building, Manor Road, Oxford, OX1 3UQ}

\author{Stefan Thurner}
\email{Corresponding author, e-mail: stefan.thurner@meduniwien.ac.at}
\affiliation{Complexity Science Hub, Metternichgasse 8, 1030 Wien}
\affiliation{Supply Chain Intelligence Institute Austria, Metternichgasse 8, 1030 Wien}
\affiliation{Medical University of Vienna, Spitalgasse 23, 1090 Vienna}
\affiliation{Santa Fe Institute, Santa Fe, 1399 Hyde Park Rd, NM 75791, USA}

\begin{abstract}
While many national and international climate policies clearly outline decarbonization targets and the timelines for achieving them, there is a notable lack of effort to objectively monitor progress. A significant share of the transition from fossil fuels to low-carbon energy will be borne by industry and the economy, requiring both the decarbonization of the electricity sector and the electrification of industrial processes. But how quickly are firms adopting low-carbon electricity? Using a unique dataset on Hungary’s national supply chain network, we analyze the energy portfolios of 25,000 firms, covering more than 75\% of gas, 70\% of electricity, and 50\% of oil consumption between 2020 and 2024. This enables us to objectively measure the trends of decarbonization efforts at the firm level. Although almost half of firms have increased their share of low-carbon electricity, more than half have reduced it. Extrapolating the observed trends, we find a transition of only 20\% of total energy consumption to low-carbon electricity by 2050. The current speed of transition in the economy is not sufficient to reach climate neutrality by 2050. However, if firms would adopt the same efforts as the decarbonization frontrunners in their industry, a low-carbon share of up to 70\% could be reached, putting climate targets within reach. We examine several firm characteristics that differentiate transitioning from non-transitioning firms. Our results are consistent with a ‘lock-in’ effect, whereby firms with a high share of fossil fuel costs relative to revenue are less likely to transition.
\end{abstract}

\maketitle

\section*{Introduction}

The global transition from fossil fuels to low-carbon energy sources is crucial for achieving international climate goals \cite{rogelj_mitigation_2018}. The energy transition has two major components: decarbonizing the electricity sector and electrifying appliances and industrial processes \cite{mercure_reframing_2021, pathak2022technical}. For some industries and processes, alternative strategies such as biomass, hydrogen, or carbon capture and storage/usage (CCS/CCU) might be more viable options for greenhouse gas mitigation \cite{bataille_review_2018, davis_net-zero_2018, rissman_technologies_2020, gailani_assessing_2024}. However, there is widespread agreement that direct electrification is the primary pathway for most end-use sectors and processes \cite{gerres_review_2019, pathak2022technical, international_energy_agency_iea_world_2024, international_renewable_energy_agency_irena_world_2024}. There is an extensive body of literature that analyzes the decarbonization of the electricity sector across different scales—global, regional, and national—using energy system models to project the rollout of low-carbon energy technologies \cite{pfenninger_energy_2014, riahi_shared_2017, capros_energy-system_2019, breyer_history_2022}. Recently, advances in electricity sector decarbonization have been observed in countries and regions around the world driven by falling prices for renewable energy technologies such as photovoltaic (PV) installations, wind power, and grid-scale batteries \cite{luderer_impact_2021, way_empirically_2022, international_energy_agency_electricity_2024}. Although many options for the electrification of industrial processes have been discussed \cite{williams_technology_2012, lechtenbohmer_decarbonising_2016, madeddu_co2_2020, world_economic_forum_net-zero_2024}, comprehensive studies on the actual adoption of these technologies remain scarce or focused on specific industries \cite{clementi_determinants_2023}.

\begin{figure*}[ht]
\centering
\includegraphics[width=13.0cm]{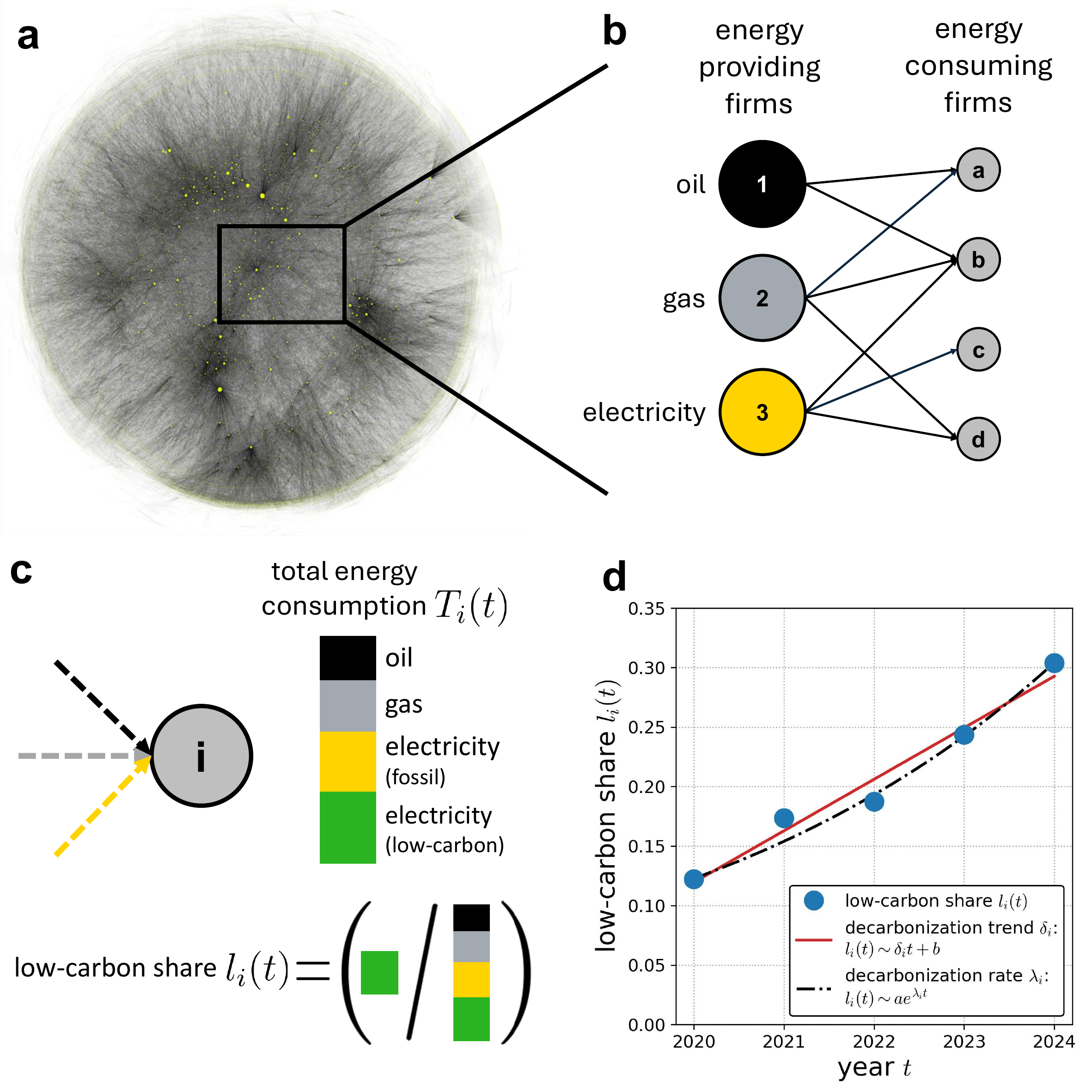}
\caption{Measuring the pace of decarbonization by reconstructing the energy mix of individual firms. (a) The supply chain network of Hungary aggregated over one year, where nodes represent firms and links represent supply relations between them. Node size corresponds to the out-strength of each firm. (b) A schematic micro-level view of the supply chain network, distinguishing between energy providing firms (left) and energy consuming firms (right). Energy providers supply electricity, gas, and oil to energy consuming firms. For every firm, annual energy consumption of each energy carrier is calculated by converting the respective payments (as reconstructed from VAT data) into kilowatt-hours using semi-annual energy prices. (c) Detailed view of the energy consumption of firm, $i$, in year, $t$. Detailed view of energy consumption for firm $i$ in year $t$. Low-carbon electricity consumption, $L_i(t)$, equals electricity consumption, $E_i(t)$, multiplied by the low-carbon share of the Hungarian electricity mix, $u(t)$ (see Methods). The firm’s low-carbon share, $l_i(t)$, is then the ratio of $L_i(t)$ to total energy consumption, $T_i(t)$. (d) Low-carbon share, $l_i(t)$, of firm $i$ calculated for every year, $t$, in the observation period. The transition toward low-carbon energy use is quantified from $l_i(t)$ by the decarbonization trend, $\delta_i$ (linear regression), and the decarbonization rate, $\lambda_i$ (exponential fit).
}
\label{Fig.:conceptual}
\end{figure*}


The literature on electrification can be divided into two dominant streams. The first examines sector-level energy consumption trends and electrification options from a top-down perspective \cite{buhler_comparative_2019, lopez_towards_2023, fraunhofer_isi_direct_2024}. Although effective in assessing industrial sectors as a whole, these studies offer limited insight into individual firms, not accounting for firm- and plant-level differences in the adoption of electrification technologies. The second stream consists of case studies on specific firms or small-scale analyses focused on energy-intensive industries regulated by emissions trading schemes or other policy initiatives \cite{stenqvist_energy_2012, wiertzema_bottomup_2020, clementi_determinants_2023, dragomir_empirical_2023}. Sector-level research is typically enabled by readily available input-output tables, which are often augmented with environmental data, known as environmentally extended input-output (EEIO) tables. These allow for an analysis of sectoral energy consumption and electrification patterns \cite{lenzen_mapping_2012, wood_global_2014}. However, this coarse sectoral view limits our understanding of the underlying dynamics, as firms within the same industry sector do not necessarily use the same technologies or follow the same adoption strategies \cite{acemoglu_advanced_2023, acemoglu_automation_2024}. Emerging research on firm-level supply chain networks highlights significant heterogeneity between firms even within fine-grained industry sector classifications, including differences in their input and output structure \cite{diem_estimating_2024}, their exposure to climate transition risks \cite{tabachova_climate_2025}, or their systemic role in the context of climate policy \cite{stangl_firm-level_2024}. Although case studies provide insights to firm-specific behaviors, barriers, and opportunities, they lack generalizability, offering only fragmented perspectives on how firms approach electrification. Some studies have used larger panels of firm- or plant level data to investigate the role of electrification for energy efficiency or productivity, but a focus on the broader energy transition is lacking \cite{boyd_estimating_2000, jung_electrification_2014}.

For the first time, we use comprehensive firm-level supply chain network data to reconstruct the energy portfolios of a large sample of firms, demonstrating how this increasingly available data can be used to analyze the dynamics of the energy transition comprehensively at the firm level. In doing so, we bridge the gap between top-down sector-level analyses and bottom-up studies of individual firm strategies, providing a comprehensive firm-level perspective on the energy transition. Firm-level supply chain network data based on value-added tax (VAT) or electronic invoicing records has recently become available in a growing number of jurisdictions \cite{pichler_building_2023}, making it possible to apply the method introduced here to monitor the ongoing energy transition in other countries and regions. Here, we analyze consecutive semi-annual snapshots of the Hungarian supply chain network, obtained from VAT data covering the years 2020 to 2024, to reconstruct the annual energy consumption portfolios of 25,231 firms. The Hungarian supply chain network has previously been studied with a focus on structural characteristics \cite{borsos_unfolding_2020}, the systemic relevance of individual firms within the supply chain \cite{diem_quantifying_2022}, the role of supply chain disruptions in amplifying financial systemic risk \cite{tabachova_estimating_2024}, and its temporal dynamics \cite{reisch_rewiring}. A recent review on the state of firm-level supply chain network research is provided by \cite{bacilieri_firm-level_2023}.

\begin{figure*}[ht] 
\centering
\includegraphics[width=13.0cm]{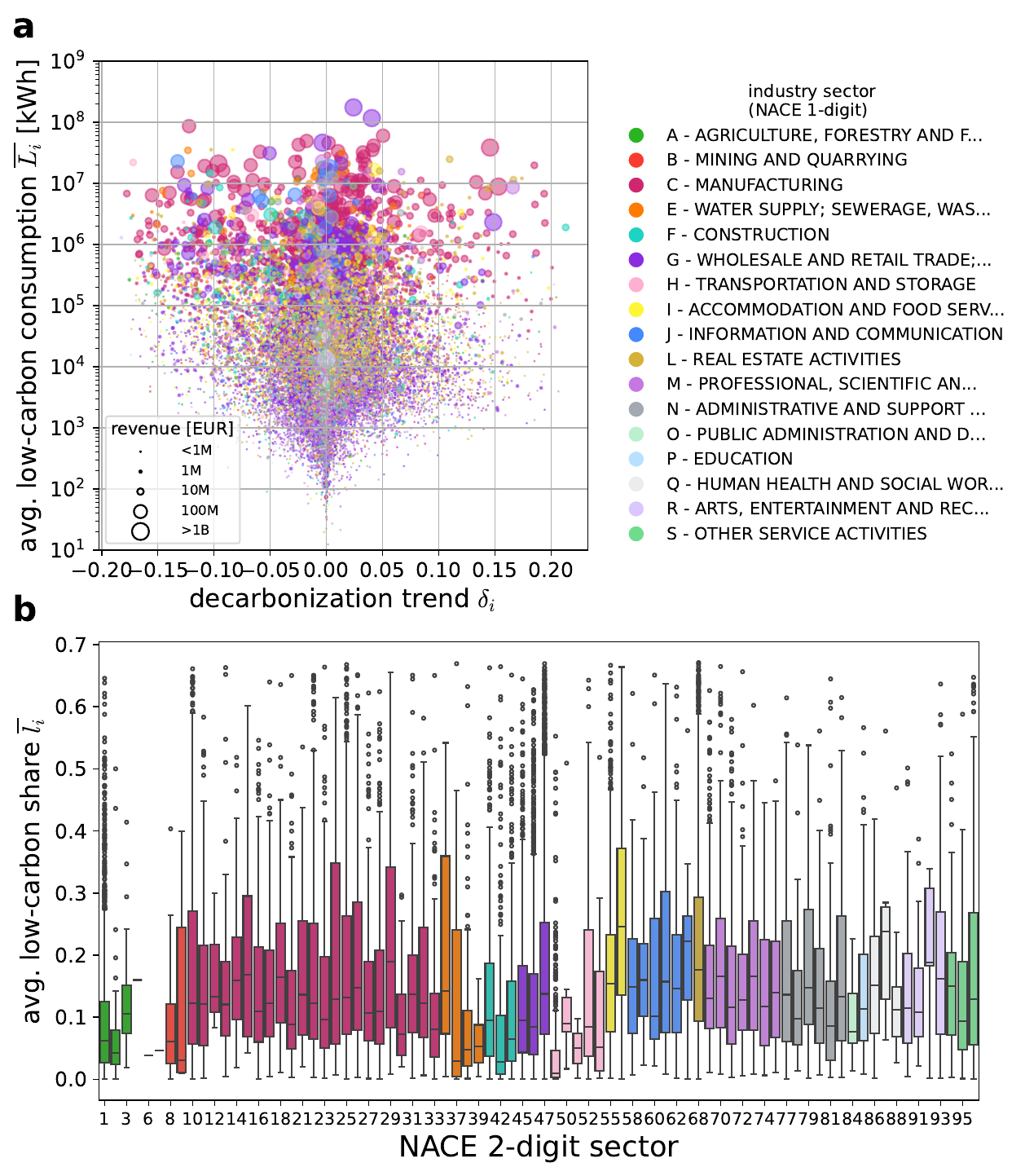} 
\caption{Low-carbon electricity consumption and decarbonization trends. (a) Scatter plot of average low-carbon consumption, $\overline{L_i}$ vs. decarbonization trend, $\delta_i$. Every dot is a firm, color indicates its NACE 1-digit industry sector, and marker sizes represents its average revenue. Higher levels of low-carbon consumption are visibly associated with larger revenue. The observed variability in $\delta_i$ highlights significant heterogeneity within and across sectors, underscoring the importance of the analysis at the firm-level. (b) Box plots of the average low-carbon share, $\overline{l_i}$, in the NACE 2-digit industry sectors. While median low-carbon shares are generally low across sectors, individual firms with high $\overline{l_i}$ are present in every sector, indicating a significant within-sector heterogeneity.}
\label{Fig.:heterogeneity}
\end{figure*}

We use the Hungarian supply chain network to identify energy-providing firms —those supplying electricity, gas, or oil products— based on their NACE 4-digit industry classification \cite{statistik_austria_systematik_2008}, as well as energy-consuming firms, as illustrated in Fig.\ref{Fig.:conceptual}a,b. We obtain a sample of 25,231 firms with continuous time series data on electricity, gas, and oil consumption, as well as revenue, and number of employees, spanning from 2020 to 2024; for details, see the Methods section. This sample covers 75\% of gas, 70\% of electricity, and 50\% of oil consumption between 2020 and 2024 relative to the total firm population. The remaining energy share not covered corresponds to firms with inconsistent time series data, which were excluded for robustness (see Methods and Supplementary Discussion). By applying semi-annual energy prices for electricity, gas, and oil, we convert monetary transactions observed in the supply chain network into kilowatt-hours of energy consumption. Low-carbon electricity consumption, $L_i(t)$, for each firm, $i$, is determined by multiplying its electricity consumption, $E_i(t)$, with the low-carbon share of Hungary's annual electricity mix, $u(t)$. Note that nuclear energy is included as a low-carbon source of electricity, in line with its classification by both the IPCC and the Hungarian government that has set a target for 90\% low-carbon electricity by 2030 \cite{pathak2022technical, ministry_for_innovation_and_technology_national_2020}. A firm’s low-carbon share, $l_i(t)$, is then calculated as the ratio of its low-carbon electricity consumption, $L_i(t)$, to its total energy consumption, $T_i(t)$, within a given year, $t$, as shown schematically in Fig.\ref{Fig.:conceptual}c. To quantify the pace of the energy transition at the firm level, we employ two approaches. First, we fit a linear function to $l_i(t)$ across the observation period to derive the \textit{decarbonization trend}, $\delta_i$, of every firm, $i$. Second, we fit an exponential function to $l_i(t)$ to estimate the \textit{decarbonization rate}, $\lambda_i$ of $i$. These indicators for the pace of the energy transition can be interpreted as two distinct patterns of technological change: an \textit{incremental mode} (linear trend) that assumes steady and gradual progress, and a \textit{disruptive mode} (exponential rate) that reflects more rapid transitions, such as those driven by capital stock renewal. Both modes have been observed and analyzed in previous studies on technology adoption \cite{geels_technological_2002, trianni_framework_2014}. With these measures, we are able to address the following questions: How heterogeneous are firms, both within and across industry sectors in their adoption of low-carbon electricity? What characteristics distinguish transitioning firms—--those with positive decarbonization trends, $\delta_i$ , and rates, $\lambda_i$,---from non-transitioning firms? And, finally, are the current trends sufficient to achieve an energy transition that aligns with international climate targets?

\section*{Results}

\textbf{Firm-level heterogeneity of energy consumption.} The analysis of firm-level energy portfolios shows significant heterogeneity in both, low-carbon electricityconsumption, and decarbonization trends, across industry sectors as shown in Fig. \ref{Fig.:heterogeneity}.

The scatter plot in Fig. \ref{Fig.:heterogeneity}a shows average low-carbon electricity consumption, $\overline{L_i}$, on the $y$-axis (logarithmic scale) versus the decarbonization trend, $\delta_i$ on the $x$-axis (linear scale) for every firm, $i$. Every marker represents one firm, colors indicate the NACE 1-digit industry category, the size represents its average revenue $\overline{R_i}$. We exclude firms from the NACE category 'D - Electricity, gas, steam and air conditioning supply' that act as energy providers, as our analysis focuses on the energy consumption of end users; see Method section for details. A complete list of the NACE 1-digit industry categories and their descriptions can be found in Supplementary Tab. V. The plot shows a wide range of decarbonization trends, with firms of all sizes, low-carbon electricity consumption levels, and sector affiliations exhibiting both, positive and negative decarbonization trends, $\delta_i$. The majority of firms show decarbonization trends centered around zero, meaning that their low-carbon share, $l_i$, remained relatively stable over the observation period. The top consumers of low-carbon electricity are primarily found in the sector 'C - Manufacturing', along with firms in the sector 'G - Wholesale and retail trade; repair of motor vehicles and motorcycles'. The plot also highlights considerable variability in decarbonization trends across both, sectors, and individual firms. While low-carbon electricity consumption and decarbonization trends span nearly the entire space, a considerable gap exists in the bottom-right corner of Fig.\ref{Fig.:heterogeneity}a, where small electricity consumers with high decarbonization rates are practically absent. This suggests that predominantly larger firms and those with higher electricity consumption dominate the progress in increasing their low-carbon share. The majority of small consumers (defined as those with low-carbon consumption, $\overline{L_i} < 10^3$ kWh) exhibit negative decarbonization trends, with 1,308 firms showing a decline of their low-carbon share compared to 762 firms displaying positive trends.

Figure \ref{Fig.:heterogeneity}b shows the distribution of the low-carbon share, $\overline{l_i}$, for the NACE 2-digit industry sectors. While median values are generally low, notable within-sector heterogeneity exists, along with variation between NACE 2-digit sectors within the same NACE 1-digit category. Manufacturing sectors, which are among the largest consumers of low-carbon electricity (as shown in Fig.\ref{Fig.:heterogeneity}a), exhibit higher median low-carbon shares. Service sectors such as 'J - Information and communication' and 'M - Professional, scientific, and technical activities', located higher up in the NACE industry classification, also show higher median values. Across all sectors, some firms approach a low-carbon share of 0.7, indicating near-exclusive use of low-carbon electricity. The upper bound for the low-carbon share is 0.7378, representing the proportion of low-carbon electricity sources in the Hungarian electricity mix in  2024 (see Methods section for more details). Although some firms have made significant progress in increasing their low-carbon share, the majority still show relatively modest levels. 

\begin{figure*}
\centering
\includegraphics[width=13cm]{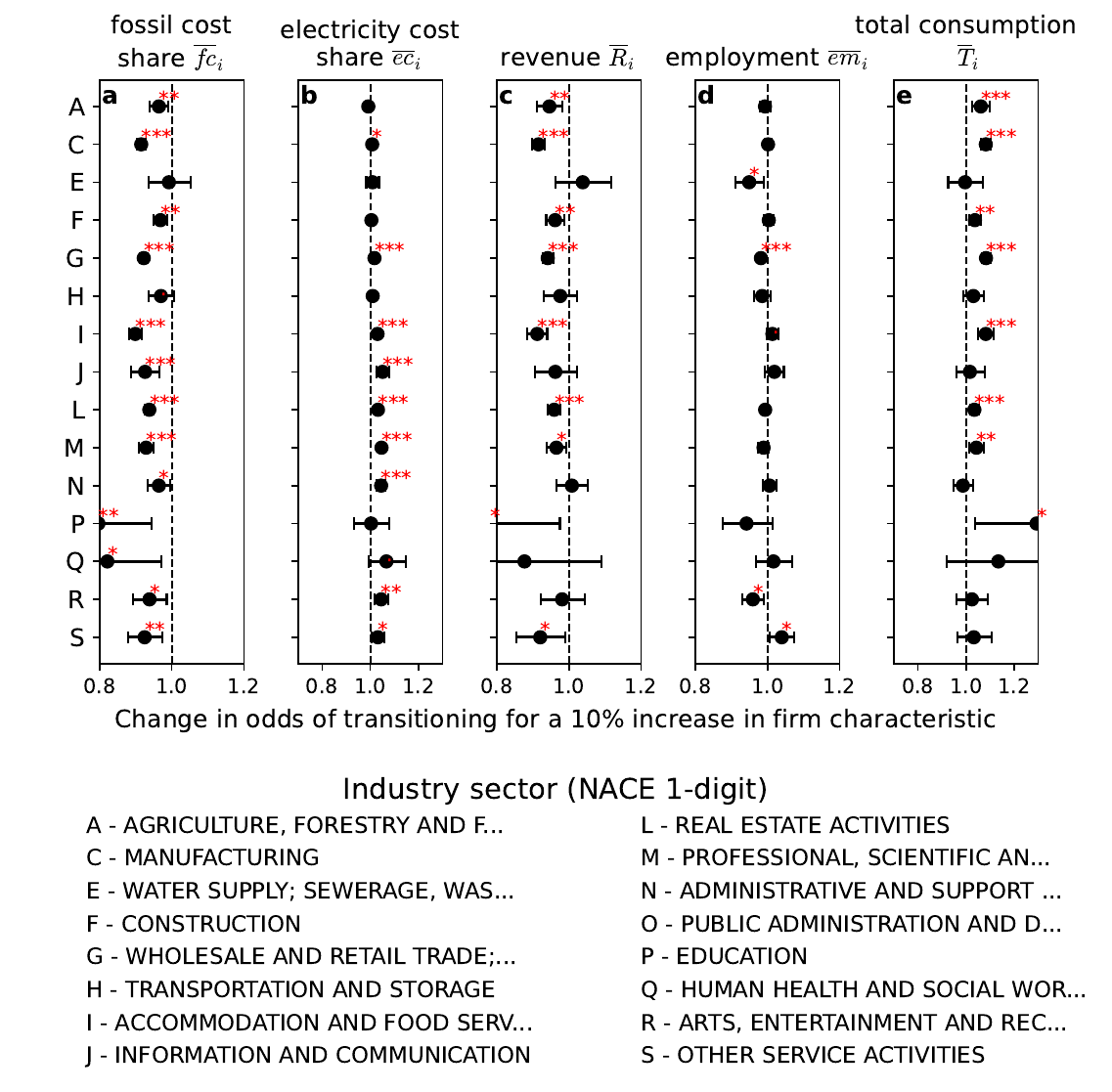} 
\caption{Differences between transitioning ($\delta_i > 0$ and $\lambda_i > 0$) and non-transitioning ($\delta_i < 0$ or $\lambda_i < 0$) firms with respect to other firm-level characteristics. The forest plot shows the the adjusted odds ratios (AORs) from multivariate logistic regressions for each NACE 1-digit sector, analyzing the relationship between firm transition status and firm characteristics. Each point represents the AOR for a 10\% increase in the corresponding characteristic, calculated as $e^{\beta \cdot 0.1}$, with horizontal bars indicating 95\% confidence intervals. The vertical dashed line marks an odds ratio of one, corresponding to no effect. Characteristics include average fossil cost share, $\overline{fc_i}$, average electricity cost share ($\overline{ec_i}$), average revenue, $\overline{R_i}$), average number employment, $\overline{em}_i$, and total consumption, $\overline{T_i}$. Statistical significance levels are denoted by superscript stars: *** if $p$-value $<$ 0.001, ** if $p$-value $<$ 0.01, * if $p$-value $<$ 0.05 and . if $p$-value $<$ 0.1. Note that we excluded sector 'B – Mining and Quarrying' from the regression analysis because too few observations were available to estimate the coefficients.
}
\label{Fig.:difference_transition}
\end{figure*}


\textbf{Characteristics of transitioning firms.} 
To determine which firms tend to transition, Fig. \ref{Fig.:difference_transition} shows the relation of several firm characteristics with their transition status: transitioning, ($\delta_i > 0$ and $\lambda_i > 0$), and not transitioning ($\delta_i < 0$ or $\lambda_i < 0$). More specifically, Fig. \ref{Fig.:difference_transition} presents a forest plot of the Adjusted Odds Ratios (AOR) from a multivariate logistic regression analysis for the following firm characteristics for each NACE 1-digit sector: (a) average fossil energy cost share of revenue, $\overline{fc}_i$, (b) average electricity cost share of revenue, $\overline{ec}_i$, (c) average revenue, $\overline{R}_i$, (d) average employment, $\overline{em}_i$, and (e) average total energy consumption, $\overline{T}_i$. Note that we exclude sector 'B – Mining and Quarrying' from the regression analysis because too few observations were available to estimate the coefficients. The AOR represents the change in odds of transitioning for a 10\% increase in the respective firm characteristic. Figure \ref{Fig.:difference_transition}a shows that a higher fossil cost share is associated with a significantly lower likelihood of transitioning across NACE 1-digit sectors. In contrast, a higher electricity cost share is associated with a greater likelihood of transitioning, as shown in Fig.~\ref{Fig.:difference_transition}b. Larger firms in terms of revenue tend to exhibit a lower likelihood of transitioning as shown in Fig.\ref{Fig.:difference_transition}c, whereas higher total energy consumption is associated with a greater likelihood of transitioning, as shown in Fig.\ref{Fig.:difference_transition}e. The number of employees does not appear to have a significant association with the likelihood of transitioning across most NACE 1-digit sectors, as shown in Fig.~\ref{Fig.:difference_transition}d. Table \ref{Tab.:logistic_regression} reports the estimated adjusted odds ratios together with their 95\% confidence intervals. The effects appear strongest for the fossil cost share and total consumption. The adjusted odds ratios for a 10\% increase in fossil energy cost share show a significant decrease in the likelihood of transitioning for most sectors, supporting the hypothesis that high fossil energy costs are associated with non-transitioning firms. The adjusted odds ratios for a 10\% increase in electricity cost share show a consistent increase in the likelihood of transitioning, supporting the hypothesis that higher electricity costs are more common among transitioning firms. While higher overall consumption seems to be associated with a higher likelihood to transition away from fossil fuels, higher revenue seems to decrease the likelihood of being a transitioning. This may indicate that firms with higher revenues can afford to continue paying rising fossil energy prices rather than making potentially risky investments to change their capital stock. These results are significant across most NACE 1-digit sectors and may indicate that a firm’s energy cost structure plays a central role in determining whether it transitions to low-carbon electricity consumption. The observed higher electricity costs for transitioning firms and higher fossil energy costs for non-transitioning firms could be explained by a 'lock-in effect', where firms are constrained by their current energy technologies. The upfront investments required to switch to low-carbon electricity may further deter firms from transitioning.

\begin{table*}
\centering
\begin{tabular}{cccccc}
\toprule
\makecell{industry Sector\\(NACE 1-digit)} & fossil cost
share $\overline{fc}_i$ & electricity cost
share $\overline{ec}_i$ & revenue $\overline{R}_i$ & employment $\overline{em}_i$ & total consumption
 $\overline{T}_i$ \\
\midrule
A & $0.964^{**} \, [0.939, 0.991]$ & $0.992^{} \, [0.981, 1.002]$ & $0.946^{**} \, [0.912, 0.981]$ & $0.993^{} \, [0.978, 1.009]$ & $1.062^{***} \, [1.025, 1.100]$ \\
C & $0.915^{***} \, [0.902, 0.929]$ & $1.008^{*} \, [1.001, 1.014]$ & $0.916^{***} \, [0.898, 0.934]$ & $1.002^{} \, [0.993, 1.010]$ & $1.082^{***} \, [1.062, 1.103]$ \\
E & $0.992^{} \, [0.936, 1.052]$ & $1.009^{} \, [0.983, 1.036]$ & $1.038^{} \, [0.964, 1.118]$ & $0.949^{*} \, [0.910, 0.990]$ & $0.996^{} \, [0.926, 1.071]$ \\
F & $0.969^{**} \, [0.950, 0.988]$ & $1.004^{} \, [0.996, 1.013]$ & $0.962^{**} \, [0.937, 0.987]$ & $1.004^{} \, [0.990, 1.018]$ & $1.037^{**} \, [1.012, 1.063]$ \\
G & $0.922^{***} \, [0.912, 0.933]$ & $1.017^{***} \, [1.012, 1.023]$ & $0.941^{***} \, [0.927, 0.956]$ & $0.982^{***} \, [0.974, 0.990]$ & $1.083^{***} \, [1.066, 1.101]$ \\
H & $0.970^{.} \, [0.936, 1.006]$ & $1.010^{} \, [0.997, 1.023]$ & $0.976^{} \, [0.931, 1.024]$ & $0.985^{} \, [0.962, 1.009]$ & $1.030^{} \, [0.987, 1.076]$ \\
I & $0.899^{***} \, [0.881, 0.917]$ & $1.030^{***} \, [1.018, 1.042]$ & $0.912^{***} \, [0.885, 0.940]$ & $1.014^{.} \, [0.998, 1.030]$ & $1.082^{***} \, [1.051, 1.114]$ \\
J & $0.926^{***} \, [0.888, 0.966]$ & $1.051^{***} \, [1.025, 1.078]$ & $0.962^{} \, [0.906, 1.022]$ & $1.019^{} \, [0.993, 1.046]$ & $1.016^{} \, [0.959, 1.077]$ \\
L & $0.938^{***} \, [0.925, 0.951]$ & $1.032^{***} \, [1.023, 1.041]$ & $0.959^{***} \, [0.942, 0.977]$ & $0.994^{} \, [0.984, 1.003]$ & $1.035^{***} \, [1.017, 1.053]$ \\
M & $0.929^{***} \, [0.909, 0.949]$ & $1.047^{***} \, [1.034, 1.059]$ & $0.965^{*} \, [0.939, 0.993]$ & $0.989^{} \, [0.975, 1.004]$ & $1.043^{**} \, [1.013, 1.074]$ \\
N & $0.964^{*} \, [0.934, 0.995]$ & $1.045^{***} \, [1.027, 1.063]$ & $1.008^{} \, [0.965, 1.053]$ & $1.006^{} \, [0.988, 1.024]$ & $0.987^{} \, [0.947, 1.029]$ \\
P & $0.797^{**} \, [0.671, 0.945]$ & $1.003^{} \, [0.932, 1.080]$ & $0.782^{*} \, [0.627, 0.975]$ & $0.942^{} \, [0.876, 1.013]$ & $1.293^{*} \, [1.036, 1.614]$ \\
Q & $0.821^{*} \, [0.694, 0.972]$ & $1.067^{.} \, [0.993, 1.146]$ & $0.877^{} \, [0.705, 1.091]$ & $1.017^{} \, [0.968, 1.068]$ & $1.134^{} \, [0.919, 1.399]$ \\
R & $0.938^{*} \, [0.893, 0.986]$ & $1.045^{**} \, [1.017, 1.074]$ & $0.981^{} \, [0.922, 1.044]$ & $0.959^{*} \, [0.930, 0.990]$ & $1.025^{} \, [0.962, 1.091]$ \\
S & $0.925^{**} \, [0.878, 0.974]$ & $1.031^{*} \, [1.004, 1.059]$ & $0.921^{*} \, [0.856, 0.991]$ & $1.040^{*} \, [1.006, 1.074]$ & $1.032^{} \, [0.963, 1.106]$ \\
\bottomrule
\end{tabular}
\caption{ Adjusted odds ratios (AORs) from multivariate logistic regressions for each NACE 1-digit sector analyzing the relationship between firm transition status and firm characteristics averaged over the observation period. Reported values are AORs for a 10\% increase in each characteristic, calculated as $e^{\beta \cdot 0.1}$, with 95\% confidence intervals in brackets. Characteristics include average fossil cost share, $\overline{fc_i}$, average electricity cost share $\overline{ec_i}$, average revenue, $\overline{R_i}$, average employment, $\overline{em_i}$, and average total consumption, $\overline{T_i}$. Statistical significance levels are denoted by stars: *** if $p$-value $<$ 0.001, ** if $p$-value $<$ 0.01, * if $p$-value $<$ 0.05 and . if $p$-value $<$ 0.1. Note that we excluded sector 'B – Mining and Quarrying' from the regression analysis because too few observations were available to estimate the coefficients.
}
\label{Tab.:logistic_regression}
\end{table*}

\begin{figure*}[ht]
\centering
\includegraphics[width=13.0cm]{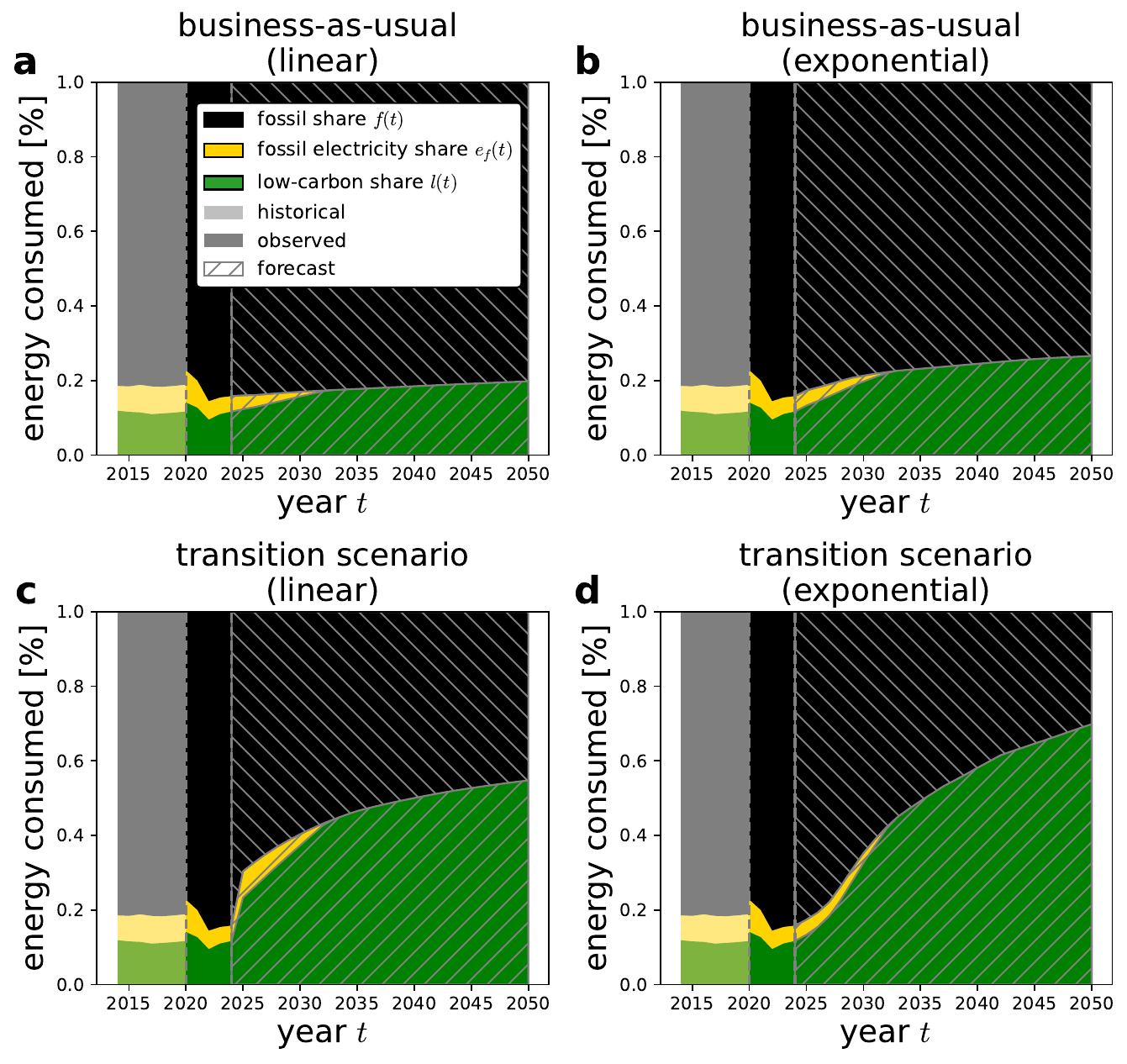}  
\caption{Scenarios of aggregate energy consumption by fossil share, $f_(t)$, fossil electricity share, $e_f(t)$, and low-carbon share $l(t)$ based on observed firm-level trends. Here, decarbonization trends and rates are determined by the interaction between firms' electrification trends and rates with forecasts of the low-carbon share of Hungary’s electricity mix $u(t)$ (see Methods and Supplementary Methods). (a) Business-as-usual (linear): The observed decarbonization trends, $\delta_i$, are extrapolated for all firms.  
(b) Business-as-usual (exponential): The observed decarbonization rates, $\lambda_i$, are extrapolated for all firms.  
(c) Transition scenario (linear): Firms with negative decarbonization trends are assigned positive decarbonization trends based on the closest firm within the same NACE 4-digit sector, in terms of revenue and number of employees; see Methods section. The aggregate low-carbon and fossil shares are then extrapolated based on these new decarbonization trends.  
(d) Transition scenario (exponential): Firms with negative decarbonization rates, $\lambda_i$, are assigned positive decarbonization rates from the closest firm within the same NACE 4-digit sector, based on revenue and number of employees; see Methods. The aggregate low-carbon and fossil shares are then extrapolated according to these new decarbonization rates.
Table \ref{Tab.:scenarios_values} presents the forecast low-carbon and fossil shares for 2020, 2030, 2040, and 2050 across for all scenarios.}
\label{Fig.:scenarios}
\end{figure*}


\begin{table*}[ht]
\centering
\setlength{\tabcolsep}{8pt} 
\renewcommand{\arraystretch}{1.2} 
\begin{tabular}{l|*{4}{c c|}}
 & \multicolumn{2}{c|}{\textbf{\makecell{business-as-usual\\(linear)}}} 
 & \multicolumn{2}{c|}{\textbf{\makecell{business-as-usual\\(exponential)}}} 
 & \multicolumn{2}{c|}{\textbf{\makecell{transition\\(linear)}}} 
 & \multicolumn{2}{c}{\textbf{\makecell{transition\\(exponential)}}} \\
\hline
{\rm year} & $l(t)$ & $f(t)$ & $l(t)$ & $f(t)$ & $l(t)$ & $f(t)$ & $l(t)$ & $f(t)$ \\
\hline
2020 & 0.141 & 0.772 & 0.141 & 0.772 & 0.141 & 0.772 & 0.141 & 0.772 \\
2030 & 0.157 & 0.830 & 0.197 & 0.787 & 0.373 & 0.597 & 0.326 & 0.647 \\
2040 & 0.185 & 0.815 & 0.245 & 0.755 & 0.500 & 0.500 & 0.582 & 0.418 \\
2050 & 0.198 & 0.802 & 0.266 & 0.734 & 0.548 & 0.452 & 0.699 & 0.301 \\
\end{tabular}
\caption{Forecasts for low-carbon share, $l(t)$, and fossil share, $f(t)$, of aggregated energy consumption for the years 2020, 2030, 2040, and 2050 under the linear and exponential business-as-usual scenarios and the linear and exponential transition scenarios, as shown in Fig. \ref{Fig.:scenarios}. Note that values $l(t)$ and $f(t)$ for 2020 are the same, since all scenarios start from the same observed low-carbon and fossil shares.} \label{Tab.:scenarios_values}
\end{table*}


\textbf{Energy transition scenarios.} To examine whether the current pace of the energy transition aligns with international climate goals, Fig.\ref{Fig.:scenarios} presents scenarios of future aggregated energy consumption for all firms in the sample. Specifically, it displays four scenarios illustrating how the energy share between fossil electricity (yellow), low-carbon electricity (green) and fossil energy (black) may evolve until 2050, based on the observed decarbonization trends, $\delta_i$, and decarbonization rates, $\lambda_i$; see Methods section for details. All panels show the aggregated fossil share, $f(t)$, fossil electricity share, $e_f(t)$, and low-carbon share, $l(t)$ of all firms in the sample during the period 2020--2024 as solid area plots. For reference, Hungary's historical shares, as derived from final energy consumption tables are displayed for the years 2014--2019 as shaded area plots with reduced opacity. While not directly comparable to the low-carbon shares estimated from the aggregated firm sample, these historical values provide a context for Hungary’s energy transition progress. Forecasted aggregated low-carbon shares, $l_(t)$, and corresponding fossil shares, $f_(t)$, are depicted as hatched area plots. For the forecasted values, we assume that firms continue to consume their average total energy, $\overline{T_i}$ estimated for the period 2020-–2024. The electricity mix of Hungary is assumed to gradually decarbonize until the early 2030s, aligning with Hungary’s goal of achieving 90\% low-carbon electricity by 2030; For details, see Methods and Supplementary Methods).

Figure \ref{Fig.:scenarios}a depicts the business-as-usual scenario, based on observed linear decarbonization trends, $\delta_i$. Firms with positive trends increase their share of low-carbon electricity, while firms with negative trends reduce it over time. The plot shows that the system quickly saturates, with $l_(t)$ reaching only  $0.198$ by 2050; see Table \ref{Tab.:scenarios_values}. This suggests that no substantial transition occurs if the current linear trends persist.

Figure \ref{Fig.:scenarios}b illustrates the business-as-usual scenario, based on observed exponential decarbonization rates, $\lambda_i$. Firms with positive rates significantly increase their share of low-carbon electricity, while those with negative rates decrease it over time. Initially, the overall low-carbon share, $l_(t)$, rises but plateaus around 2030, as most firms with positive rates fully decarbonize by then. By 2050, the low-carbon share reaches a value of only $0.266$. This indicates that even with exponential adoption of electrification technologies, the system-wide transition to low-carbon electricity would fall short of meeting international climate goals. Firms currently increasing their fossil energy shares must reverse this trend, as explored in the transition scenarios in Fig.\ref{Fig.:scenarios}c and Fig.\ref{Fig.:scenarios}d.

Figure \ref{Fig.:scenarios}c shows a transition scenario, where every firm with a negative decarbonization trend, $\delta_i$, adopts a positive trend derived from an industry peer within the same NACE 4-digit sector, matched by average revenue and employment. This means that every firm adopts the positive electrification strategies of its peers, resulting in all firms having positive decarbonization trends, $\delta_i$ from $t = 2024$ onward. In this optimistic scenario, the forecasted low-carbon share, $l_(t)$, rises sharply until the mid-2030s, as Hungary’s electricity mix becomes fully decarbonized. Beyond that point $l_(t)$ continues to grow steadily, reaching a share of $0.548$ by 2050; see Table \ref{Tab.:scenarios_values}. However, even with industry-wide adoption of positive linear decarbonization trends, $\delta_i$, the energy transition on the firm level remains insufficient to achieve the rapid emission reductions required to meet climate targets , i.e., net-zero emissions by 2050 \cite{rogelj_mitigation_2018}.
Figure \ref{Fig.:scenarios}d presents the 'best-case' scenario, where every firm with a negative exponential decarbonization rate, $\lambda_i$, adopts a positive rate from $t = 2024$ onward from an industry peer within the same NACE 4-digit sector, matched by average revenue and employment. This scenario assumes that all firms adopt the best-case decarbonization strategies of their peers, resulting in universal positive decarbonization rates. Here, the forecasted low-carbon share, $l_(t)$, increases rapidly until the early 2030s, when Hungary’s electricity mix is  forecasted to be fully decarbonized. Beyond this point $l_(t)$ continues to climb steeply, reaching a share of $0.699$ by 2050. This suggests that in a best-case scenario, where all firms pivot decisively towards decarbonization, the energy transition at the firm level could become consistent with achieving the Paris climate goals, especially given that electrification is typically associated with significant gains in energy efficiency gains \cite{pathak2022technical}. Remaining fossil fuel consumption could additionally be compensated by other means than direct electrification, such as biomass, hydrogen and synthetic fuels, or carbon capture and storage/usage, which are not accounted for in this analysis. We provide an uncertainty analysis of the presented energy scenarios in the Supplementary Discussion and Supplementary Fig. 15.

\FloatBarrier

\section*{Discussion}

Achieving climate neutrality in line with international targets is critically dependent on the success of the energy transition, much of which must be borne by industry and the wider economy. Although existing research has largely concentrated on the decarbonization of the electricity sector, comprehensive evidence on how firms themselves are adopting low-carbon electricity remains scarce. For the first time, we use detailed firm-level supply chain network data to reconstruct the energy portfolios of 25,231 of the most relevant energy-consuming firms in Hungary. This approach provides an unprecedented level of granularity, allowing us to analyze the dynamics and pace of the energy transition at the firm level and to uncover patterns that remain hidden in aggregate statistics.

We find remarkable heterogeneity both within and between industry sectors concerning low-carbon electricity consumption, the share of low-carbon electricity in the overall energy mix, and the decarbonization speeds of firms -- defined as the speed of increase in the low-carbon electricity share between 2020 and 2024. While almost half (49.6\%) of the firms in the sample showed positive decarbonization trends, more than half of the firms actually reduced their low-carbon electricity share. This is a concerning indication that the energy transition is far from assured, and a huge fraction of firms continues to rely on fossil energy sources.

We find that transitioning and non-transitioning firms display distinct and consistent characteristics across most NACE 1-digit sectors. Transitioning firms are more likely to exhibit higher electricity cost shares, while non-transitioning firms tend to have higher fossil cost shares. These results suggest that energy cost structures play a central role in shaping transition behavior. Our regression analysis further indicates that greater total energy consumption is associated with a higher likelihood of transitioning, which may suggest that more energy-intensive firms face stronger pressures to reduce fossil dependence. By contrast, higher revenues are associated with a lower likelihood of transitioning, possibly reflecting that wealthier firms are better able to absorb rising fossil fuel costs rather than commit to potentially risky investments in new capital stock. Employment levels, in turn, show little association with the likelihood of transitioning. Taken together, these findings may be explained by what has been described in the literature as 'lock-in effect': firms’ existing energy technologies can constrain their flexibility, making fossil-reliant firms less likely to shift away and electricity-reliant firms more likely to move toward low-carbon energy consumption. The substantial upfront investments required to replace fossil-based capital may reinforce such dynamics, potentially slowing the transition even when long-term economic or environmental incentives exist. This investment barrier has also been highlighted in previous studies \cite{pathak2022technical, international_energy_agency_iea_world_2024, international_renewable_energy_agency_irena_world_2024}

To assess whether current trends align with international climate goals, we simulate a set of energy scenarios based on observed decarbonization trajectories and the potential for shifts in firms’ strategies. The analysis indicates that current energy consumption dynamics are insufficient to achieve meaningful decarbonization in the near future, with low-carbon shares projected to reach only 20--26\% by 2050. This outcome is largely attributable to the fact that many firms continue to expand their fossil fuel use, thereby offsetting the progress made by transitioning firms. At the same time, we identify untapped potential to accelerate the transition. In virtually all fine-grained NACE 4-digit sector, we find firms that are already increasing their share of low-carbon electricity, suggesting that no fundamental technological barriers hinder the transition. To illustrate this potential, we construct and analyze two transition scenarios in which firms adopt the decarbonization trajectories of the frontrunners in their respective industries. Universal adoption of frontrunners’ linear trends would raise the low-carbon share to 55\% by 2050, whereas a best-case scenario in which firms adopt the frontrunners’ exponential growth rates would lift it to 70\%. These transition scenarios demonstrate that substantial progress is feasible if firms converge toward the performance of industry leaders. Note that these energy scenarios are primarily illustrative, showing both the insufficiency of current trends and the potential for acceleration. They do not represent a comprehensive assessment of Hungary’s overall energy consumption.

While our findings reveal robust correlations between firms’ energy cost structures, size, and their transition behavior, our empirical strategy does not allow us to establish causal relationships. The policy implications outlined below should therefore be interpreted with appropriate caution. Future research could build on our results by identifying the underlying mechanisms, potentially through quasi-experimental designs or other causal inference approaches (see Supplementary Discussion). Nevertheless, policymakers have a broad set of instruments to facilitate effective transition pathways, and our findings point to several potential avenues for policy design. One option is lowering the investment barriers that could prevent firms from shifting away from fossil-based technologies. Targeted support, such as subsidies for electrification technologies (e.g., heat pumps) or state-backed loans with favorable conditions, can help firms manage the high upfront costs of replacing existing capital. Once these investments are made it is plausible that firms continue their low-carbon paths, reinforcing progress over time.
At the same time, the finding that high-revenue firms are less likely to transition could point to the need for instruments that make continued fossil fuel use less attractive. Stronger carbon pricing is one broad option. In addition, since our study provides a tool to identify frontrunners and laggards within fine-grained industry segments, policymakers could design more targeted measures, such as sectoral emission standards or performance-based incentives that reward greener firms through mechanisms like tax breaks or preferential access to public procurement. Encouragingly, we find a positive correlation between total energy consumption and transition behavior, which is particularly important given the leverage of large consumers in reducing emissions. Large energy consumers that demonstrate a credible decarbonization pathway could likewise benefit from favorable treatment.

Our study has several limitations. The data analyzed covers the period from 2020 to 2024. Prior to 2020, the quality of firm-level data is less reliable due to changes in the thresholds for reporting VAT transactions. Specifically, firms were required to report VAT transactions only when they reached a certain transaction value threshold, which decreased over time. After 2020, this threshold was removed entirely, as described in \cite{bacilieri_firm-level_2023}. Only then a complete reconstruction of the energy portfolios of firms becomes possible. Although 2020-2024 allows one to obtain trends in the firm-level energy transition, it is a relatively short time span. Nevertheless, this period is particularly relevant, as electrification and renewable energy technologies have become more affordable, and climate policies, such as those in the EU Green Deal introduced in late 2019, gained increasing momentum among policymakers and industry. In this respect, the 2020-2024 period represents a crucial phase for observing the onset of the energy transition.

Fluctuations in energy prices, particularly for electricity and gas during the energy crisis, may have influenced our results. In 2022, for example, we observed a decline in the overall low-carbon share in the scenarios shown in Fig.~\ref{Fig.:scenarios}. This decrease may be linked to the sharp rise in electricity prices during the crisis, which appears to be only partly reflected in the expenditure patterns of our firm sample. While our sample reasonably reflects relative oil and gas consumption, it appears to under-represent electricity use and over-represent gas use compared to official statistics, possibly because firms procure electricity from additional, unobserved sources such as own production or the spot market (see Supplementary Discussion). Overall, our methodology is more reliable in periods of relatively stable energy prices, as indicated by the stable aggregate shares for 2023 and 2024. Further details on how energy price data were incorporated are provided in Methods and Supplementary Methods.

Our study faces another limitation related to the conversion of energy use into costs using EUROSTAT price data stratified by consumption bands. This approach masks heterogeneity in effective prices within each band, arising from differences in load profiles, contractual arrangements, on-site generation, and access to renewable procurement. For instance, organizations with predominantly daytime loads, such as universities, logistics hubs, or service-sector firms, may face lower effective electricity prices due to greater opportunities for solar self-generation, while energy-intensive industrial firms operating around the clock may face higher effective prices. These differences also imply varying opportunities for decarbonization, since daytime-oriented consumers are better positioned to integrate self-generation or time-sensitive procurement options. As a result, we may underestimate the low-carbon share of firms with daytime-oriented load profiles and overestimate it for firms with more even, industrial load profiles. If this bias is systematic, the general trends in our fitted models are unlikely to change substantially, although the intercept may shift. Therefore, the forecasted aggregate low-carbon shares in the scenario analysis (Fig.\ref{Fig.:scenarios}; Tab.~\ref{Tab.:scenarios_values}) may be somewhat overstated, given the high energy use of industrial consumers. Our results should therefore be interpreted with this limitation in mind, and future work could incorporate more granular pricing data or load-profile information to better capture these firm-level cost differences.

A related limitation is that we do not observe energy generation activities carried out directly on firms’ premises, such as photovoltaic (PV) installations. Our analysis is restricted to electricity purchases from other companies and therefore omits behind-the-meter PV generation that firms self-consume. Calculations provided in the Supplementary Discussion indicate that commercially self-consumed PV electricity increased from roughly 0.6~TWh in 2020 to about 2.5~TWh in 2024, corresponding to around 27\% of total national PV output in recent years and covering an estimated 2.3\% of total commercial electricity demand in 2020 and about 8.8\% in 2024. Although this growth is substantial, the overall magnitude remains modest relative to total commercial consumption, particularly for large industrial users. However, the impact is likely heterogeneous across sectors: service-sector firms may exhibit comparatively higher self-consumption shares relative to their overall electricity use, whereas large industrial firms typically consume much larger volumes, making the contribution of PV self-consumption relatively small. As a consequence, our derived low-carbon electricity shares for large industrial consumers are unlikely to be materially affected. This may not hold for firms that have invested in large-scale PV systems dedicated for self-consumption \cite{MAVIR2025}, but because we cannot identify these firms, this remains a limitation of our analysis. By contrast, the low-carbon electricity shares $l_i(t)$ for smaller service-sector firms may be understated and should therefore also be interpreted with caution. Future work could address this limitation by integrating data on solar PV installations if such information becomes available.

Energy providers typically sell both, gas and electricity but are categorized under only one NACE code, either for the distribution of gas or electricity. To address this issue, we limited our analysis to firms with connections to both gas and electricity providers, assuming they would purchase gas and electricity from separate providers. This restriction resulted in a sample of 25,231 firms, which still accounts for a substantial share of total energy consumption in Hungary (25\% of total final energy consumption in 2023, including transport and residential consumption: 16.6\% of oil, 40.0\% of gas, and 17.1\% of electricity), making it a reasonable representation of the broader firm population. Further details on this selection process are found in the Methods section. A detailed discussion of the resulting firm sample is presented in the the Supplementary Discussion.

Oil products are highly heterogeneous, ranging from fuels such as gasoline and diesel to feedstocks like naphtha. Since we do not have product-level information but only observe monetary transactions in the supply chain between firms, we must make assumptions about which oil products firms consume in order to convert monetary values into kilowatt-hours via prices. Specifically, we assumed that firms consume oil in the form of diesel and gasoline and applied the weighted average price of these fuels in Hungary to convert expenditures into kilowatt-hours. This provides a reasonable estimate of the energy consumed through oil products, as diesel and gasoline are by far the most widely consumed oil products in Hungary, with the exception of the chemical industries where naphtha is the most prevalent. For more details, see the Methods and Supplementary Methods.

Our focus is on electrification as a pathway to low-carbon energy, which is viable for most sectors but less applicable in industries such as chemicals or cement, where alternative strategies are needed. Future research could examine these alternatives if detailed product-level data on firm-to-firm transactions becomes available, enabling a broader assessment of decarbonization pathways beyond electrification and a more comprehensive understanding of firm-level transitions.

The generalisability of our findings beyond Hungary should be treated with caution. Certain features of the Hungarian economy—such as its continued reliance on fossil fuels, the presence of energy-intensive industries, the expansion of nuclear power, and the rapid uptake of solar PV—are also characteristic of other Central and Eastern European countries. While the quantitative results remain context-specific, some qualitative insights are likely to extend more broadly, including the structural role of energy costs in shaping transition behaviour and the finding that nearly all fine-grained industry sectors contain both frontrunners rapidly adopting low-carbon electricity and laggards continuing to expand fossil fuel use.

Future research could extend our analysis by examining how suppliers and customers influence a firm’s transition behavior. An exploratory assessment of this aspect, indicates that firms’ transition dynamics are more closely aligned with those of their customers than with their suppliers (see Supplementary Discussion).

To conclude, our study shows that across all industrial sectors, both frontrunners and laggards coexist—even within fine-grained NACE 4-digit classifications grouping firms with highly similar activities. This demonstrates that a technical pathway to low-carbon electricity exists in virtually every industry. At the same time, the fact that many firms continue to move in the opposite direction highlights the urgent need for stronger policies to oversee, incentivize, and steer the transition at the firm level. By leveraging firm-level supply chain network data, we provide a new lens for analyzing decarbonization dynamics, establishing a foundation for more detailed investigations of firm strategies. As such data becomes increasingly available across regions \cite{pichler_building_2023}, our methodology offers a scalable and transferable tool for monitoring and comparing national energy transitions and for identifying best practices that can accelerate progress.

\section*{Methods}
\subsection*{Reconstructing annual firm energy use from the Hungarian supply chain network}
\label{methods_firm-level_network}
This study relies on the reconstructed firm-level supply chain network of the Hungarian economy, derived from value-added tax (VAT) transaction data collected by the Hungarian National Tax and Customs Administration. Since 2014, this data, made available through the Central Bank of Hungary, has enabled the identification of supply relationships between firms, facilitating the creation of detailed snapshots of the Hungarian supply chain network at the firm level. More information on this dataset can be found in previous studies \cite{borsos_unfolding_2020, diem_quantifying_2022}. The dataset anonymizes the companies but provides information such as revenue, number of employees, and industry sector, enabling analysis based on company size and activities. The coverage of the dataset has evolved over time due to changes in reporting thresholds for VAT transactions. Between early 2015 and mid-2018, only transactions with a cumulative tax content exceeding 1 million Hungarian Forint (HUF) within a reporting period (monthly, quarterly, or annually) were recorded. From the third quarter of 2018 to mid-2020, the threshold was lowered to 100,000 HUF and applied to individual transactions, significantly increasing the visibility of firms and their supply relationships. However, firms whose transactions consistently fell below this new threshold were excluded. Since the third quarter of 2020, all inter-firm invoices must be reported, eliminating thresholds entirely and providing a comprehensive view of supply chain relationships. This is also discussed in detail in \cite{bacilieri_firm-level_2023}. Larger firms in Hungary are required to report their VAT on a monthly basis, which allows a fine-grained resolution of the monetary transactions between firms. Only smaller firms are required to report on an annual basis.

\subsubsection*{Identifying energy providers}
For this study, we use semi-annual snapshots of the Hungarian firm-level supply chain network from 2020 to 2024 \st{2023}, as this period offers the most consistent and comprehensive time-series data following the removal of reporting thresholds. To identify energy providers within the network, we categorize firms based on their NACE 4-digit industry affiliations. Electricity providers are identified as firms classified under one or more of the following categories: 'D35.1 - Electric power generation, transmission, and distribution,' 'D35.1.1 - Production of electricity,' 'D35.1.2 - Transmission of electricity,' 'D35.1.3 - Distribution of electricity,' and 'D35.1.4 - Trade of electricity.' Gas providers are identified as firms in the categories 'D35.2.1 - Manufacture of gas,' 'D35.2.2 - Distribution of gaseous fuels through mains,' and 'D35.2.3 - Trade of gas through mains.' Oil providers are identified as firms classified under 'B6.1.0 - Extraction of crude petroleum,' 'C19.2.0 - Manufacture of refined petroleum products,' 'G47.3.0 - Retail sale of automotive fuel in specialized stores,' and 'G46.7.1 - Wholesale of solid, liquid, and gaseous fuels and related products.' It is important to note that multiple NACE 4-digit categories are included for each type of energy provider (electricity, gas, and oil) to account for the limitations of industry classifications. Firms' NACE classifications do not always capture the full scope of their activities. For instance, a firm might simultaneously transport and sell gas or refine and sell oil, yet only be classified under a single category. Moreover, energy companies often operate as large entities consisting of multiple sub-companies with different industry classifications. To ensure comprehensive coverage of energy consumption, we include all relevant categories associated with these activities. Note that coal usage is not covered in this study due to the absence of a distinct NACE code specifically related to coal distribution. Additionally, coal in Hungary is primarily used for electricity production, which is already addressed in the electricity mix of Hungary. Beyond that, coal consumption is limited to a small number of firms, mostly in the steel and pulp and paper industries \cite{IEA2022}.

\subsubsection*{Firm sample construction}
We aggregate the monetary inputs from companies identified as providers of electricity, gas, and oil---based on the industry classifications detailed above---on a semi-annual basis over the observation period from 2020 to 2024. This aggregation enables us to estimate the total amount of electricity, gas, and oil purchased by each firm semi-annually, allowing us to track their energy consumption trends over time. Firms with an annual reporting requirement are kept in the sample. To ensure data quality, we apply several restrictions to our sample. Since energy providers often supply both gas and electricity, we retain only firms that simultaneously purchase electricity and gas from companies classified under electricity and gas provider categories, based on their NACE 4-digit sector affiliation, for each year of the observation period. This ensures we account for firms with distinct electricity and gas purchases. Firms classified under any NACE 4-digit industry related to energy provision---whether electricity, gas, or oil---are excluded to focus on end-users of energy, rather than energy suppliers. Additionally, firms from the financial sector, classified under the NACE 1-digit category "K - Financial and insurance activities" are excluded, as they may act as energy brokers without directly consuming energy themselves. Furthermore, firms from the NACE 4-digit category 'H52.2.1 - Service activities incidental to land transportation' have been excluded from the analysis, as firms from this category might be involved in liquefaction of gas for transportation purposes which likely means their purchases are not linked to consumption of gas directly. We exclude firms participating in the European Union Emissions Trading System (ETS), as their energy costs are affected by additional expenditures for carbon credits, which cannot be accounted for in our analysis of firm characteristics. Our dataset is anonymized, allowing us to identify whether a company participates in the ETS but not to determine which specific firms are involved. Estimating their energy-related carbon credit costs would require individual treatment that is not feasible within the scope of this study. For this reason, we adopt a conservative approach and exclude ETS firms from the sample, amounting to 201 firms in total. Firms without a NACE classification in the dataset, and those lacking revenue data for any observed year, are also excluded. We further exclude an outlier firm whose gas consumption increased anomalously by three orders of magnitude from one year to the next. After applying these criteria, the final sample consists of 25,231 firms for which continuous semi-annual time-series data on electricity, gas, oil consumption, and revenue are available. Although this filtering approach may exclude firms that exclusively use gas or electricity, it is necessary to maintain the consistency and reliability of the resulting firm sample. Since energy providers often supply both gas and electricity, and we cannot distinguish between these energy types in the monetary transactions within our dataset, the assumption of separate providers for electricity and gas ensures the robustness of the analysis. Supplementary Discussion gives details on the firm sample coverage and its comparison with sectoral energy consumption data from Hungary’s energy balance \cite{mekh_energy_balance_2025}.

\subsubsection*{Conversion of monetary inputs into energy consumption via energy prices}
We use semi-annual energy prices to convert the monetary inputs for electricity, gas, and oil into kilowatt-hours of energy consumed by each firm in the sample. For electricity and gas, we rely on energy price data for non-household users in Hungary from EUROSTAT, which includes all taxes and levies \cite{eurostat_gas_2024, eurostat_electricity_2024}. This data is collected from electricity and gas providers who report the prices paid by their customers across different consumption bands (seven for electricity and six for gas). We determine the appropriate consumption band for each firm by converting the energy consumption thresholds into monetary units using the respective energy prices. We then assign each firm to the corresponding band for each half-year based on its expenditures. For firms with an annual VAT reporting requirement, we derive annual energy prices by taking the average of the semi-annual energy prices. This applies to between 1,000 and 5,000 firms in any given year in our sample. After applying the semi-annual energy prices, we aggregate oil, gas, and electricity consumption to the annual level for further processing to avoid seasonality effects.
Supplementary Figs. 2 and 3 show the price trends for electricity and gas. Over the observation period, electricity and gas prices have fluctuated significantly, particularly during the energy crisis of 2022, when prices surged. They subsequently decreased again in 2023. It is also important to note that price evolution varies across different consumer groups: larger firms typically experienced earlier price increases and reductions, while smaller firms saw these changes later. This discrepancy may be attributed to larger firms hedging their energy costs through the energy futures market, while smaller firms often have fixed contracts with one or more energy providers. 

\textbf{Government support during the energy crisis.} In 2022, the Hungarian government introduced a support scheme aimed at energy-intensive small- and medium-sized enterprises (SMEs), covering 50\% of the increased electricity and gas costs. This scheme impacted an estimated 10,000 companies \cite{about_hungary2023}. In 2023, the government also implemented a price cap on electricity, benefiting around 5,000 companies in sectors such as manufacturing, accommodation, and warehousing/transport \cite{reuters2023}. These market interventions do not affect our energy consumption estimates, as the SME support scheme functioned as a reimbursement, and the electricity price cap applied directly to the energy bills of companies. Consequently, the prices reported by EUROSTAT still represent the actual prices paid by companies to their energy providers and, therefore, the corresponding kilowatt-hours consumed.

\subsubsection*{Estimating oil consumption based on fuel prices}
To estimate oil consumption, we assume that firms primarily use oil in the form of fuels and employ fuel price trends as a proxy to convert observed oil expenditures into energy units. Data from the National Detailed Energy Balance, provided by the Hungarian Energy and Public Utility Regulatory Authority (MEKH), confirms that diesel and gasoline are the dominant forms of oil product consumption \cite{mekh_energy_balance_2025}. Supplementary Fig. 4 shows the distribution of oil product consumption for 2023, derived from final consumption data in the National Detailed Energy Balance. Diesel accounts for nearly half of oil product consumption in 2023, followed by gasoline at 20\%. Naphtha represents 14\% of total consumption; however, according to the National Detailed Energy Balance, the chemical and petrochemical industries are the sole consumers of naphtha. Therefore, the assumption that firms primarily consume oil in the form of fuels is reasonable for all industry sectors, except for the chemical industry, which also consumes oil in the form of naphtha. Since naphtha is generally cheaper than diesel or gasoline, our method may underestimate oil product consumption in the chemical sector. 
Fuel price data is obtained from historical trends in the Weekly Oil Bulletin provided by the EU \cite{eu_oil_bulletin}. To determine a single representative price for oil products, we calculate a weighted average of gasoline and diesel prices in Hungary, using weights based on the relative consumption of these fuels, also derived from the Weekly Oil Bulletin data. This approach implicitly assumes that firms consume these fuels in similar proportions. On average, diesel represents 74\% of fuel consumption, while gasoline accounts for 26\% during the observation period from 2020 to 2024. This weighted average price allows us to convert monetary expenditures on oil products into kilowatt-hours consumed. Supplementary Fig. 5 depicts the evolution of fuel prices for gasoline, diesel, and their weighted average in Hungary from 2018 to 2024. Fuel prices have risen significantly since 2021, reflecting the broader energy crisis in Europe.

\subsubsection*{Calculating the low-carbon share $l_i(t)$}
To determine the low-carbon electricity consumption $L_i(t)$ for a firm $i$ in a given year, we use the low-carbon share of Hungary's annual electricity mix, as provided by the online platform Ember \cite{ember_electricity_2023}. Ember reports annual data on clean and fossil electricity generation in terawatt-hours, excluding imports and exports. We compute the low-carbon share of Hungary's annual electricity mix, $u(t)$, by dividing the electricity generated from low-carbon sources (including hydro, PV, wind, bioenergy, and nuclear) by the total electricity produced. The low-carbon electricity consumption of a firm in a given year is calculated as
\begin{equation}
L_i(t) = E_i(t) \cdot u(t) \quad , 
\end{equation}
where $E_i(t)$ represents the total electricity consumption of firm $i$. The low-carbon share of a firm's total energy consumption is then given by
\begin{equation}
l_i(t) = \frac{L_i(t)}{T_i(t)} \quad , 
\end{equation}
where $T_i(t)$ denotes the total energy consumption of firm $i$ in year $t$. This approach allows us to estimate the low-carbon share $l_i(t)$ for each firm in our sample over the period from 2020 to 2024, enabling us to analyze trends and assess the pace at which firms are increasing the low-carbon share of their energy mix. We therefore assume that all firms source from the same electricity mix and thus do not account for individual efforts to procure clean electricity. However, given that Hungary’s electricity market is highly concentrated, with one firm group holding a leading position in the market, this assumption may be considered reasonable for the average firm sourcing from the most prevalent provider (see also \cite{IEA2022}).

\subsection*{Measuring the speed of the energy transition at the firm level}
To measure the pace of the energy transition for individual firms, we use two distinct models of technology adoption. The first approach assumes that the decarbonization process follows a gradual, steady trend, while the second approach models the transition as a potentially faster, more disruptive process.

\textbf{Linear model.} In the first model, we fit a linear function to the low-carbon share, $l_i(t)$, for each firm $i$ over the observation period. This allows us to calculate the \textit{decarbonization trend}, $\delta_i$, which represents the gradual pace of decarbonization. The linear equation is
\begin{equation}
l_i(t) = \alpha_i + \delta_i \cdot t \quad , 
\end{equation}
where $\alpha_i$ is the initial low-carbon share and $\delta_i$ is the slope of the trend. 

Instead of ordinary least squares (OLS), we employ a robust regression estimator based on the Huber loss function \cite{Huber1964}, which down-weights the influence of large residuals by combining quadratic and linear loss.
\begin{equation}
\rho(r) = 
\begin{cases}
\frac{1}{2} r^2 & \text{if } |r| \leq k \\
k |r| - \frac{1}{2} k^2 & \text{if } |r| > k
\end{cases}
\end{equation}
for residuals $r = l_i(t) - \alpha_i - \delta_i \cdot t$, where $k$ is a tuning constant. We set $k$ to the default value used in the Python \texttt{statsmodels} implementation, $k = 1.345$ \cite{seabold_statsmodels_2010}. Small residuals are treated quadratically (as in OLS), while large residuals receive linear weighting, reducing the influence of outliers. The optimal values of $\delta_i$ and $\alpha_i$ are thus obtained by solving
\begin{equation}
\label{eq:regression_linear}
(\hat{\alpha}_i, \hat{\delta}_i) = \arg\min_{\alpha_i, \delta_i} \sum_{t=2020}^{2024} \rho\!\left( l_i(t) - \alpha_i - \delta_i \cdot t \right) \quad .
\end{equation}

\textbf{Exponential model.}  
The second model assumes that the low-carbon share, $l_i(t)$ follows an exponential trajectory, with decarbonization characterized by a potentially faster growth process:
\begin{equation}
l_i(t) = \beta_i \cdot e^{\lambda_i \cdot t} \quad ,
\end{equation}
where $\beta_i > 0$ is the initial level and $\lambda_i$ is the exponential growth rate.  

For estimation, we take logarithms, which transforms the exponential specification into a linear model in log-space:
\begin{equation}
\ln \big( l_i(t) \big) = \gamma_i + \lambda_i \cdot t \quad ,
\end{equation}
where $\gamma_i = \ln(\beta_i)$. This allows us to estimate $\lambda_i$ as the slope of a regression of $\ln(l_i(t))$ on time. We employ the same robust regression procedure as in the linear model, now applied to the log-transformed data. For residuals
\begin{equation}
r = \ln \big( l_i(t) \big) - \gamma_i - \lambda_i \cdot t \, ,
\end{equation}
the optimal values $(\hat{\gamma}_i, \hat{\lambda}_i)$ are obtained by solving
\begin{equation}
\label{eq:regression_exponential}
(\hat{\gamma}_i, \hat{\lambda}_i) = \arg\min_{\gamma_i, \lambda_i} \sum_{t=2020}^{2024} \rho \!\left( \ln \big( l_i(t) \big) - \gamma_i - \lambda_i \cdot t \right) \, ,
\end{equation}
where $\rho(\cdot)$ denotes the Huber loss function defined above. The original parameter of interest $\beta_i$ can then be recovered as $\hat{\beta}_i = e^{\hat{\gamma}_i}$. Supplementary Methods and Supplementary Fig. 6 demonstrate that robust estimation provides more reliable results than OLS by reducing the influence of outliers in selected example firms.

While logistic growth models are also commonly used in studies of technological adoption, we refrain from using them here. This is because logistic growth requires fitting a three-parameter model, and with only five time points 2020-2024, such a model would not be reliable or meaningful for this analysis. Instead, the linear and exponential models are better suited for illustrating two distinct patterns of technological change: steady, incremental transitions and more rapid, disruptive shifts.

\subsection*{Quantifying the relationship between firm characteristics and transition status}

We perform multivariate logistic regression analyses to examine the relationship between firm characteristics and their transition status within each NACE 1-digit industry sector. Transitioning firms are defined as those with a positive decarbonization trend $\delta_i > 0$ and a positive exponential decarbonization rate $\lambda_i > 0$, while non-transitioning firms have negative values for either of the two indicators. Ambiguous cases, where only one of the two indicators is negative, are conservatively also classified as non-transitioning. The binary outcome variable is coded as 1 for transitioning firms and 0 otherwise.

As predictors, we use averaged firm-level variables over the observation period, applying logarithmic transformations to mitigate the influence of extreme values or skewness. The predictor set combines indicators of energy costs and firm size: average fossil energy cost share, $\overline{fc_i}$, average electricity cost share $\overline{ec_i}$, average firm revenue $\overline{R_i}$, average employment $\overline{em_i}$, and average total energy consumption $\overline{T_i}$. 

For each NACE 1-digit sector, we estimate a logistic regression of the form
\begin{equation}
\text{logit}(P(\text{trans}_i)) = \alpha + \sum_j \beta_j X_{ij} \quad ,
\end{equation}
where $P(\text{trans}_i)$ is the probability that firm $i$ transitions, and $X_{ij}$ denotes the log-transformed predictor variables. The regression coefficients $\beta_j$ capture how each firm characteristic relates to the probability of transitioning, conditional on the others.  

To aid interpretation, we compute odds ratios associated with a 10\% increase in each predictor:
\begin{equation}
\text{OR}_{1\%} = e^{\beta_j \cdot 0.1} \quad ,
\end{equation}
with corresponding confidence intervals based on robust standard errors. An odds ratio greater than 1 indicates that an increase in the predictor raises the odds of transitioning, while a value below 1 suggests the opposite. 

\subsection*{Constructing energy transition scenarios}

\subsubsection*{Decarbonization of Hungary's electricity sector}

To create scenarios for future low-carbon electricity consumption, we need to make a forecast of the evolution of low-carbon share $u_(t)$ in Hungary's national electricity mix. This share influences the low-carbon share $l_i(t)$ for each firm $i$, as it reflects the overall decarbonization progress of the country's electricity sector.
Between 2020 and 2024, Hungary's low-carbon share grew significantly, from about 60\% in 2020 to over 70\% in 2024. This increase was mainly due to the expansion of photovoltaic (PV) installations. To forecast $u(t)$ for 2025 to 2050, we use a linear regression based on the observed trend from 2020 to 2024. The resulting forecast for $u_(t)$, is consistent with Hungary's target of achieving 90\% low-carbon electricity by 2030. This would require an annual increase of 3.2 in the share of low-carbon electricity. Under this forecast, Hungary’s electricity sector will be fully decarbonized by 2033, and remain so thereafter. Supplementary Fig. 1 and Supplementary Tab. I provide further details on the observed values of $u(t)$ and the forecasted path for $u_(t)$.

\subsubsection*{Forecasting future low-carbon shares $l_{i,\text{forecast}}(t)$}

To forecast the future low-carbon share $l_i(t)$ for each firm $i$, we first need to make an assumption regarding future total energy consumption $T_i(t)$. We assume that each firm’s energy consumption will remain at the average level observed between 2020 and 2024 $\overline{T_i} = \frac{1}{5} \sum_{t=2020}^{2024} T_i(t)$.

This assumption is conservative because many electricity-powered appliances and processes are more efficient than their fuel-based counterparts \cite{pathak2022technical}. However, the potential overestimation of energy use is counterbalanced by the possibility of firm growth and capital expansion, which may increase energy consumption.

It is important to note that the observed decarbonization trends $\delta_i$ and decarbonization rates, $\lambda_i$ of firms cannot be directly used to forecast their future low-carbon shares because, $l_i(t)$ already reflects the change of Hungary’s low-carbon share, $u(t)$. Therefore, we perform separate robust regressions to estimate a \textit{linear electrification trend} $\epsilon_i$, and an \textit{exponential electrification rate} $\mu_i$ for each firm, analogous to the regressions in equations \ref{eq:regression_linear} and \ref{eq:regression_exponential}. These regressions are based on the share of electricity in firms’ energy mixes, excluding Hungary’s low-carbon share $e_i(t) = \frac{E_i(t)}{T_i(t)}$

Depending on the scenario—whether a linear or exponential adoption is assumed—the forecasted low-carbon shares for each firm $i$ in the years 2025 to 2050 are calculated using one of the following equations
\begin{equation}
\label{eq:trend_forecast}
l_{i}(t) = (\alpha_i + \epsilon_i \cdot t) \cdot u_(t) \quad , 
\end{equation}
or
\begin{equation}
\label{eq:rate_forecast}
l_{i}(t) = (\beta_i \cdot e^{\mu_i \cdot t}) \cdot u_(t) \quad .
\end{equation}
In both equations $\alpha_i$ and $\beta_i$ represent the initial values for each firm’s electrification trend or rate. The forecasted fossil share is then calculated as
\begin{equation}
f_{i}(t) = 1 - l_{i}(t)
\quad .
\end{equation}

\subsubsection*{Modeling the linear business-as-usual scenario}
In the linear business-as-usual scenario, firms continue their current electrification trends $\epsilon_i$, while Hungary’s electricity mix gradually decarbonizes. The forecasted low-carbon share for each firm $i$ in year $t$ is calculated according to equation \ref{eq:trend_forecast}. To project the total low-carbon and fossil energy shares consumed by all firms from 2025 to 2050, we first compute the low-carbon and fossil energy consumption for each firm $i$ in each year $t$
\begin{equation}
L_{i}(t) = l_{i}(t) \cdot \overline{T_i}\quad , 
\end{equation}
\begin{equation}
F_{i}(t) = f_{i}(t) \cdot \overline{T_i} \quad , 
\end{equation}
where $\overline{T_i}$ represents the firm's total energy consumption, assumed constant over time. The total low-carbon share of energy consumption across all firms is then given by
\begin{equation}
l_(t) = \sum_{i=1}^{n} \frac{L_{i}(t)}{\sum_{j=1}^{n} \overline{T_j}} \quad , 
\end{equation}
where $n$ is the number of firms. Finally, the total fossil energy share is determined as $f_(t) = 1 - l_(t)$.

\subsubsection*{Modeling the exponential business-as-usual scenario}
In the exponential business-as-usual scenario, firms continue to follow their current electrification rates $\mu_i$, while Hungary's electricity mix gradually decarbonizes. The forecasted low-carbon share for each firm $i$ in year $t$ is computed using equation \ref{eq:rate_forecast}. The total low-carbon and fossil shares of energy consumption across all firms are then determined using the same methodology as detailed above.

\subsubsection*{Constructing the linear transition scenario}
In the transition scenario, all firms are required to decarbonize by adopting a positive electrification trend $\epsilon_i$. If a firm $i$ does not exhibit a positive electrification trend, it is assigned a trend based on a matched firm $j$ with similar characteristics. The matching process follows these steps:

\begin{enumerate}
    \item Identify all firms within the same NACE 4-digit sector as firm $i$.
    \item Select a firm $j$ from this group that most closely matches firm $i$ in terms of average revenue $\overline{R_i}$ and average number of employees $\overline{em_i}$  using a Nearest Neighbor search. Revenue data is available from 2020 to 2023 and employment data is available from 2020 to 2022.
    \item If no suitable match is found at the NACE 4-digit level (i.e., no firm in the same sector has a positive electrification trend), the search expands to the NACE 2-digit sector, following the same matching procedure.
\end{enumerate}

Through this procedure, each firm is assigned a positive decarbonization trend $\mu_i$. The forecasted low-carbon share for each firm $i$ in year $t$ is then computed using equation \ref{eq:trend_forecast}. The total low-carbon and fossil shares are calculated as previously described.

\subsubsection*{Constructing the exponential transition scenario}
In the exponential transition scenario, all firms are required to decarbonize rapidly by adopting a positive electrification rate $\mu_i$. For any firm $i$ that does not follow a positive electrification rate, it is paired with a firm $j$ exhibiting the most similar characteristics. The pairing process follows the same procedure outlined earlier. As a result, every firm is assigned a positive electrification rate $\mu_i$. The forecasted low-carbon share for each firm $i$ in year $t$ is then calculated according to equation \ref{eq:rate_forecast}, and the total low-carbon and fossil shares are computed in the same manner as described above.

\section*{Data Availability}
Data on financial transactions between Hungarian value-added tax paying firms that support the findings of this study are available at the National Bank of Hungary but restrictions apply to the availability of these data, which were used under license for the current study through directly accessing the servers of the National Bank of Hungary, and so are not publicly available. Requests for collaborations can be addressed to \url{olahzs@mnb.hu}. 

\section*{Code Availability}
The code supporting the analyses, including the aggregation of purchases related to energy inputs, the estimation of decarbonization trends and rates, the multivariate logistic regressions of firm characteristics, and the construction of the energy scenarios, is available at \url{https://github.com/jo-stangl/using_firm-level_supply_chain_networks_to_measure_the_speed_of_the_energy_transition}.

\bibliographystyle{naturemag}
\bibliography{main}

@article{lenzen_mapping_2012,
  title = {Mapping the {Structure} of the {World} {Economy}},
  volume = {46},
  issn = {0013-936X, 1520-5851},
  doi = {10.1021/es300171x},
  urldate = {2023-10-18},
  journal = {Environmental Science \& Technology},
  author = {Lenzen, Manfred and Kanemoto, Keiichiro and Moran, Daniel and Geschke, Arne},
  month = aug,
  year = {2012},
  pages = {8374--8381}
}

@article{wood_global_2014,
	title = {Global {Sustainability} {Accounting}—{Developing} {EXIOBASE} for {Multi}-{Regional} {Footprint} {Analysis}},
	volume = {7},
	issn = {2071-1050},
	doi = {10.3390/su7010138},
	 
	number = {1},
	urldate = {2023-10-18},
	journal = {Sustainability},
	author = {Wood, Richard and Stadler, Konstantin and Bulavskaya, Tatyana and Lutter, Stephan and Giljum, Stefan and De Koning, Arjan and Kuenen, Jeroen and Schütz, Helmut and Acosta-Fernández, José and Usubiaga, Arkaitz and Simas, Moana and Ivanova, Olga and Weinzettel, Jan and Schmidt, Jannick and Merciai, Stefano and Tukker, Arnold},
	month = dec,
	year = {2014},
	pages = {138--163},
}

@article{mercure_reframing_2021,
	title = {Reframing incentives for climate policy action},
	volume = {6},
	issn = {2058-7546},
	doi = {10.1038/s41560-021-00934-2},
	 
	number = {12},
	urldate = {2024-09-03},
	journal = {Nature Energy},
	author = {Mercure, J.-F. and Salas, P. and Vercoulen, P. and Semieniuk, G. and Lam, A. and Pollitt, H. and Holden, P. B. and Vakilifard, N. and Chewpreecha, U. and Edwards, N. R. and Vinuales, J. E.},
	month = nov,
	year = {2021},
	pages = {1133--1143},
}

@article{diem_estimating_2024,
	title = {Estimating the loss of economic predictability from aggregating firm-level production networks},
	volume = {3},
	copyright = {https://creativecommons.org/licenses/by/4.0/},
	issn = {2752-6542},
	doi = {10.1093/pnasnexus/pgae064},
	number = {3},
	urldate = {2024-10-15},
	journal = {PNAS Nexus},
	author = {Diem, Christian and Borsos, András and Reisch, Tobias and Kertész, János and Thurner, Stefan},
	editor = {Jaworski, Taylor},
	month = feb,
	year = {2024},
	pages = {pgae064},
}

@article{stenqvist_energy_2012,
	title = {Energy efficiency in energy-intensive industries—an evaluation of the {Swedish} voluntary agreement {PFE}},
	volume = {5},
	copyright = {http://www.springer.com/tdm},
	issn = {1570-646X, 1570-6478},
	doi = {10.1007/s12053-011-9131-9},
	 
	number = {2},
	urldate = {2024-10-15},
	journal = {Energy Efficiency},
	author = {Stenqvist, Christian and Nilsson, Lars J.},
	month = may,
	year = {2012},
	pages = {225--241},
}

@article{jung_electrification_2014,
	title = {Electrification and productivity growth in {Korean} manufacturing plants},
	volume = {45},
	issn = {01409883},
	doi = {10.1016/j.eneco.2014.07.022},
	 
	urldate = {2024-10-15},
	journal = {Energy Economics},
	author = {Jung, Yonghun and Lee, Seong-Hoon},
	month = sep,
	year = {2014},
	pages = {333--339},
}

@article{diem_quantifying_2022,
	title = {Quantifying firm-level economic systemic risk from nation-wide supply networks},
	volume = {12},
	issn = {2045-2322},
	doi = {10.1038/s41598-022-11522-z},
	number = {1},
	urldate = {2024-12-03},
	journal = {Scientific Reports},
	author = {Diem, Christian and Borsos, András and Reisch, Tobias and Kertész, János and Thurner, Stefan},
	month = may,
	year = {2022},
	pages = {7719},
}

@article{stangl_firm-level_2024,
	title = {Firm-level supply chains to minimize unemployment and economic losses in rapid decarbonization scenarios},
	volume = {7},
	issn = {2398-9629},
	doi = {10.1038/s41893-024-01321-x},
	 
	number = {5},
	urldate = {2024-12-03},
	journal = {Nature Sustainability},
	author = {Stangl, Johannes and Borsos, András and Diem, Christian and Reisch, Tobias and Thurner, Stefan},
	month = apr,
	year = {2024},
	pages = {581--589},
}

@article{way_empirically_2022,
	title = {Empirically grounded technology forecasts and the energy transition},
	volume = {6},
	issn = {25424351},
	url = {https://linkinghub.elsevier.com/retrieve/pii/S254243512200410X},
	doi = {10.1016/j.joule.2022.08.009},
	 
	number = {9},
	urldate = {2024-12-03},
	journal = {Joule},
	author = {Way, Rupert and Ives, Matthew C. and Mealy, Penny and Farmer, J. Doyne},
	month = sep,
	year = {2022},
	pages = {2057--2082},
}

@techreport{bacilieri_firm-level_2023,
	type = {Working {Paper}},
	title = {Firm-level production networks: what do we (really) know?},
	url = {https://www.inet.ox.ac.uk/publications/no-2025-14-firm-level-production-networks-what-do-we-really-know},
	number = {2025-14},
	institution = {INET Oxford Working Paper Series},
	author = {Bacilieri, Andrea and Borsos, András and Astudillo-Estevez, Pablo and Lafond, François},
	year = {2025},
}

@article{tabachova_estimating_2024,
	title = {Estimating the impact of supply chain network contagion on financial stability},
	volume = {75},
	issn = {15723089},
	doi = {10.1016/j.jfs.2024.101336},
	 
	urldate = {2024-12-03},
	journal = {Journal of Financial Stability},
	author = {Tabachová, Zlata and Diem, Christian and Borsos, András and Burger, Csaba and Thurner, Stefan},
	month = dec,
	year = {2024},
	pages = {101336},
}

@techreport{borsos_unfolding_2020,
	address = {Budapest},
	type = {{MNB} {Occasional} {Papers}},
	title = {Unfolding the hidden structure of the {Hungarian} multi-layer firm network},
	copyright = {http://www.econstor.eu/dspace/Nutzungsbedingungen},
	language = {eng},
	number = {139},
	institution = {Magyar Nemzeti Bank},
	author = {Borsos, Andras and Stancsics, Martin},
	year = {2020},
}

@article{boyd_estimating_2000,
	title = {Estimating the linkage between energy efficiency and productivity},
	volume = {28},
	copyright = {https://www.elsevier.com/tdm/userlicense/1.0/},
	issn = {03014215},
	doi = {10.1016/S0301-4215(00)00016-1},
	 
	number = {5},
	urldate = {2024-12-03},
	journal = {Energy Policy},
	author = {Boyd, Gale A. and Pang, Joseph X.},
	month = may,
	year = {2000},
	pages = {289--296},
}

@book{international_renewable_energy_agency_irena_world_2024,
	address = {Abu Dhabi},
	title = {World {Energy} {Transitions} {Outlook} 2024: 1.5°{C} {Pathway}},
	isbn = {978-92-9260-634-3},
	publisher = {International Renewable Energy Agency},
	author = {{International Renewable Energy Agency (IRENA)}},
	year = {2024},
}

@techreport{international_energy_agency_iea_world_2024,
	address = {Paris},
	title = {World {Energy} {Outlook} 2024},
	url = {https://www.iea.org/reports/world-energy-outlook-2024},
	institution = {IEA},
	author = {{International Energy Agency (IEA)}},
	year = {2024},
}

@incollection{rogelj_mitigation_2018,
	title = {Mitigation {Pathways} {Compatible} with 1.5°{C} in the {Context} of {Sustainable} {Development}},
	booktitle = {Global {Warming} of 1.5°{C}. {An} {IPCC} {Special} {Report} on the {Impacts} of {Global} {Warming} of 1.5°{C} {Above} {Pre}-industrial {Levels} and {Related} {Global} {Greenhouse} {Gas} {Emission} {Pathways}, in the {Context} of {Strengthening} the {Global} {Response} to the {Threat} of {Climate} {Change}, {Sustainable} {Development}, and {Efforts} to {Eradicate} {Poverty}},
	author = {Rogelj, Joeri and Shindell, Drew and Jiang, Ke and Fifita, Sefanaia and Forster, Piers and Ginzburg, Veronika and Handa, Chandra and Kheshgi, Haroon and Kobayashi, Shinya and Kriegler, Elmar and Mundaca, Luis and Séférian, Romain and Vilariño, María Victoria},
	editor = {Masson-Delmotte, Valérie and Zhai, Philippe and Pörtner, Hans-Otto and Roberts, David and Skea, Judith and Shukla, P.R. and Pirani, Anna and Moufouma-Okia, Wanjiku and Péan, Christophe and Pidcock, Robin and Connors, Sarah and Matthews, Jay B.R. and Chen, Ying and Zhou, Xiaoye and Gomis, M.I. and Lonnoy, Emmanuel and Maycock, Trevor and Tignor, M. and Waterfield, Tom},
	year = {2018},
}

@article{lechtenbohmer_decarbonising_2016,
	title = {Decarbonising the energy intensive basic materials industry through electrification – {Implications} for future {EU} electricity demand},
	volume = {115},
	issn = {03605442},
	doi = {10.1016/j.energy.2016.07.110},
	 
	urldate = {2024-12-04},
	journal = {Energy},
	author = {Lechtenböhmer, Stefan and Nilsson, Lars J. and Åhman, Max and Schneider, Clemens},
	month = nov,
	year = {2016},
	pages = {1623--1631},
}

@book{fraunhofer_isi_direct_2024,
	title = {Direct electrification of industrial process heat. {An} assessment of technologies, potentials and future prospects for the {EU}. {Study} on behalf of {Agora} {Industry}.},
	url = {https://www.agora-industry.org/fileadmin/Projects/2023/2023-20_IND_Electrification_Industrial_Heat/A-IND_329_04_Electrification_Industrial_Heat_WEB.pdf},
	urldate = {2024-12-17},
	author = {Fraunhofer ISI},
	year = {2024},
}

@book{world_economic_forum_net-zero_2024,
	title = {Net-{Zero} {Industry} {Tracker} 2024},
	url = {https://reports.weforum.org/docs/WEF_Net_Zero_Industry_Tracker_2024.pdf},
	publisher = {World Economic Forum},
	author = {{World Economic Forum}},
	year = {2024},
}

@misc{mekh_industry_final_energy_2025,
  author = {{Hungarian Energy and Public Utility Regulatory Authority}},
  title = {8.2 Final energy use of industrial sector 2020-2023},
  year = {2025},
  url = {https://www.mekh.hu/download/f/a1/91000/8_2_Ipar_vegso_felhasznalas_reszletes_eves_2020_2023.xlsx},
  note = {Accessed September 14, 2025}
}

@article{davis_net-zero_2018,
	title = {Net-zero emissions energy systems},
	volume = {360},
	issn = {0036-8075, 1095-9203},
	doi = {10.1126/science.aas9793},
	number = {6396},
	urldate = {2024-12-17},
	journal = {Science},
	author = {Davis, Steven J. and Lewis, Nathan S. and Shaner, Matthew and Aggarwal, Sonia and Arent, Doug and Azevedo, Inês L. and Benson, Sally M. and Bradley, Thomas and Brouwer, Jack and Chiang, Yet-Ming and Clack, Christopher T. M. and Cohen, Armond and Doig, Stephen and Edmonds, Jae and Fennell, Paul and Field, Christopher B. and Hannegan, Bryan and Hodge, Bri-Mathias and Hoffert, Martin I. and Ingersoll, Eric and Jaramillo, Paulina and Lackner, Klaus S. and Mach, Katharine J. and Mastrandrea, Michael and Ogden, Joan and Peterson, Per F. and Sanchez, Daniel L. and Sperling, Daniel and Stagner, Joseph and Trancik, Jessika E. and Yang, Chi-Jen and Caldeira, Ken},
	month = jun,
	year = {2018},
	pages = {eaas9793},
}

@article{clementi_determinants_2023,
	title = {Determinants of {Renewable} {Energy} {Adoption}: {Evidence} from {Italian} {Firms}:},
	volume = {42},
	issn = {2032-5355},
	shorttitle = {Determinants of {Renewable} {Energy} {Adoption}},
	doi = {10.3917/jie.pr1.0143},
	number = {3},
	urldate = {2024-12-17},
	journal = {Journal of Innovation Economics \& Management},
	author = {Clementi, Enrico Luca and Garofalo, Giuseppe},
	month = aug,
	year = {2023},
	pages = {201--234},
}

@misc{ministry_for_innovation_and_technology_national_2020,
  title        = {National {Clean} {Development} {Strategy}},
  url          = {https://unfccc.int/sites/default/files/resource/LTS_1_Hungary_2021_EN.pdf},
  urldate      = {2024-12-18},
  publisher    = {{Ministry for Innovation and Technology}},
  author       = {{Ministry for Innovation {and} Technology}},
  year         = {2020},
  note         = {Accessed December 18, 2024},
}

@article{pichler_building_2023,
	title = {Building an alliance to map global supply networks},
	volume = {382},
	issn = {0036-8075, 1095-9203},
	doi = {10.1126/science.adi7521},
	number = {6668},
	urldate = {2024-12-18},
	journal = {Science},
	author = {Pichler, Anton and Diem, Christian and Brintrup, Alexandra and Lafond, François and Magerman, Glenn and Buiten, Gert and Choi, Thomas Y. and Carvalho, Vasco M. and Farmer, J. Doyne and Thurner, Stefan},
	month = oct,
	year = {2023},
	pages = {270--272},
}

@misc{ember_electricity_2023,
    title = {Electricity {Data} {Explorer}},
    url = {https://ember-energy.org/data/electricity-data-explorer/},
    author = {{Ember}},
    year = {2023},
    note = {Accessed August 18, 2025},
}

@misc{eurostat_gas_2024,
    title = {Gas prices for non-household consumers - bi-annual data (from 2007 onwards)},
    url = {https://ec.europa.eu/eurostat/databrowser/view/nrg_pc_203__custom_12406662/default/table?lang=en},
    author = {{Eurostat}},
    year = {2024},
    doi = {10.2908/nrg_pc_203},
    note =  {Accessed August 18, 2025},
}

@misc{eu_oil_bulletin,
    title = {Weekly Oil Bulletin Price developments 2005 onwards},
    url = {https://energy.ec.europa.eu/document/download/906e60ca-8b6a-44e7-8589-652854d2fd3f_en?filename=Weekly_Oil_Bulletin_Prices_History_maticni_4web.xlsx},
    author = {{European Commission}},
    year = {2024},
    note = {Accessed August 18, 2025},
}

@misc{eurostat_electricity_2024,
    title = {Electricity prices for non-household consumers - bi-annual data (from 2007 onwards)},
    url = {https://ec.europa.eu/eurostat/databrowser/view/nrg_pc_205__custom_9257772/default/table?lang=en},
    author = {{Eurostat}},
    year = {2024},
    doi = {10.2908/nrg_pc_205},
    note = {Accessed August 18, 2025},
}

@misc{mekh_energy_balance_2025,
  author = {{Hungarian Energy and Public Utility Regulatory Authority}},
  title = {National detailed Energy Balance - IEA format - (Annual) 2014-2024},
  year = {2025},
  url = {https://www.mekh.hu/download/2/63/b1000/7_3_orszagos_eves_IEA_tipusu_reszletes_energiamerleg_2014_2024.xlsx},
  note = {Accessed November 25, 2025}
}

@misc{Forsense2024,
    author = {{forsense}},
    title = {Megújuló energiaforrásokkal kapcsolatos trendek és vélemények Magyarországon},
    year = {2024},
    url = {https://kekbolygoalapitvany.hu/wp-content/uploads/2025/02/202412_Megujulo-energiaforrasok_KBA_2024_.pdf},
    note = {Accessed November 25, 2025}
}

@misc{EnergyCharts2025,
    author    = {{Fraunhofer Institute for Solar Energy Systems ISE}},
    title     = {Public net electricity generation in Hungary in November 2025},
    year      = {2025},
    url       = {https://energy-charts.info/charts/energy/chart.htm?l=en&c=HU},
    note      = {Accessed November 25, 2025}
}

@misc{ENTSOE2025,
    author    = {{ENTSO-E}},
    title     = {ENTSO-E Transparency Platform},
    year      = {2025},
    url       = {https://transparency.entsoe.eu/},
    note      = {Accessed November 25, 2025}
}

@article{Granger1969,
  title = {Investigating Causal Relations by Econometric Models and Cross-spectral Methods},
  volume = {37},
  ISSN = {0012-9682},
  url = {http://dx.doi.org/10.2307/1912791},
  DOI = {10.2307/1912791},
  number = {3},
  journal = {Econometrica},
  publisher = {JSTOR},
  author = {Granger,  C. W. J.},
  year = {1969},
  month = aug,
  pages = {424}
}

@article{Dumitrescu2012,
  title = {Testing for Granger non-causality in heterogeneous panels},
  volume = {29},
  ISSN = {0264-9993},
  url = {http://dx.doi.org/10.1016/j.econmod.2012.02.014},
  DOI = {10.1016/j.econmod.2012.02.014},
  number = {4},
  journal = {Economic Modelling},
  publisher = {Elsevier BV},
  author = {Dumitrescu,  Elena-Ivona and Hurlin,  Christophe},
  year = {2012},
  month = jul,
  pages = {1450–1460}
}

@misc{MEKH_1_2025,
  author = {{Hungarian Energy and Public Utility Regulatory Authority}},
  title = {4.2 Annual data for gross electricity generation 2014-2024},
  year = {2025},
  url = {https://www.mekh.hu/download/d/43/b1000/4_2_brutto_villamos_energia_termeles_eves_2014_2024.xlsx},
  note = {Accessed November 25, 2025}
}

@misc{MEKH_Methodology_2025,
  author = {{Hungarian Energy and Public Utility Regulatory Authority}},
  title = {Methodology: 4.2 Annual data for gross electricity generation},
  year = {2025},
  url = {https://www.mekh.hu/download/7/bf/01000/4_2_brutto_villamos_energia_termeles.pdf},
  note = {Accessed December 2, 2025}
}

@misc{MAVIR2025,
  author       = {{MAVIR Zrt.}},
  title        = {PV Statisztika},
  howpublished = {\url{https://mavir.hu/documents/10258/291845564/PV+STATISZTIKA_20250101-ig_v1_HU.pdf}},
  note         = {Accessed November 25, 2025},
  year         = {2025}
}

@misc{reuters2023,
  author = {Reuters},
  title = {Hungarian government to cap prices for companies' power contracts},
  year = {2023},
  publisher = {Reuters},
  url = {https://www.reuters.com/business/energy/hungarian-government-cap-prices-set-companies-power-contracts-2023-06-19/},
  note = {Accessed January 28, 2025}
}

@misc{about_hungary2023,
  author = {{About Hungary}},
  title = {Government announces support scheme for energy-intensive SMEs},
  year = {2023},
  url = {https://abouthungary.hu/news-in-brief/government-announces-support-scheme-for-energy-intensive-smes},
  note = {Accessed January 28, 2025}
}

@misc{tabachova_climate_2025,
  author        = {Zlata Tabachová and Christian Diem and Johannes Stangl and András Borsos and Stefan Thurner},
  title         = {Combined climate stress testing of supply-chain networks and the financial system with nation-wide firm-level emission estimates},
  year          = {2025},
  eprint        = {2503.10644},
  archivePrefix = {arXiv},
  primaryClass  = {q-fin.GN},
  doi           = {10.48550/arXiv.2503.10644},
  url           = {https://arxiv.org/abs/2503.10644}
}

@incollection{pathak2022technical,
  author = {M. Pathak and R. Slade and P.R. Shukla and J. Skea and R. Pichs-Madruga and D. {Ürge-Vorsatz}},
  title = {Technical Summary},
  booktitle = {Climate Change 2022: Mitigation of Climate Change. Contribution of Working Group III to the Sixth Assessment Report of the Intergovernmental Panel on Climate Change},
  editor = {P.R. Shukla and J. Skea and R. Slade and A. Al Khourdajie and R. van Diemen and D. McCollum and M. Pathak and S. Some and P. Vyas and R. Fradera and M. Belkacemi and A. Hasija and G. Lisboa and S. Luz and J. Malley},
  publisher = {Cambridge University Press},
  year = {2022},
  address = {Cambridge, UK and New York, NY, USA},
  doi = {10.1017/9781009157926.002}
}

@article{pfenninger_energy_2014,
	title = {Energy systems modeling for twenty-first century energy challenges},
	volume = {33},
	issn = {13640321},
	doi = {10.1016/j.rser.2014.02.003},
	 
	urldate = {2024-12-04},
	journal = {Renewable and Sustainable Energy Reviews},
	author = {Pfenninger, Stefan and Hawkes, Adam and Keirstead, James},
	month = may,
	year = {2014},
	pages = {74--86},
}

@article{breyer_history_2022,
	title = {On the {History} and {Future} of 100\% {Renewable} {Energy} {Systems} {Research}},
	volume = {10},
	copyright = {https://creativecommons.org/licenses/by/4.0/legalcode},
	issn = {2169-3536},
	doi = {10.1109/ACCESS.2022.3193402},
	urldate = {2024-12-04},
	journal = {IEEE Access},
	author = {Breyer, Christian and Khalili, Siavash and Bogdanov, Dmitrii and Ram, Manish and Oyewo, Ayobami Solomon and Aghahosseini, Arman and Gulagi, Ashish and Solomon, A. A. and Keiner, Dominik and Lopez, Gabriel and Ostergaard, Poul Alberg and Lund, Henrik and Mathiesen, Brian V. and Jacobson, Mark Z. and Victoria, Marta and Teske, Sven and Pregger, Thomas and Fthenakis, Vasilis and Raugei, Marco and Holttinen, Hannele and Bardi, Ugo and Hoekstra, Auke and Sovacool, Benjamin K.},
	year = {2022},
	pages = {78176--78218},
}

@article{riahi_shared_2017,
	title = {The {Shared} {Socioeconomic} {Pathways} and their energy, land use, and greenhouse gas emissions implications: {An} overview},
	volume = {42},
	issn = {09593780},
	shorttitle = {The {Shared} {Socioeconomic} {Pathways} and their energy, land use, and greenhouse gas emissions implications},
	doi = {10.1016/j.gloenvcha.2016.05.009},
	 
	urldate = {2024-12-04},
	journal = {Global Environmental Change},
	author = {Riahi, Keywan and Van Vuuren, Detlef P. and Kriegler, Elmar and Edmonds, Jae and O’Neill, Brian C. and Fujimori, Shinichiro and Bauer, Nico and Calvin, Katherine and Dellink, Rob and Fricko, Oliver and Lutz, Wolfgang and Popp, Alexander and Cuaresma, Jesus Crespo and Kc, Samir and Leimbach, Marian and Jiang, Leiwen and Kram, Tom and Rao, Shilpa and Emmerling, Johannes and Ebi, Kristie and Hasegawa, Tomoko and Havlik, Petr and Humpenöder, Florian and Da Silva, Lara Aleluia and Smith, Steve and Stehfest, Elke and Bosetti, Valentina and Eom, Jiyong and Gernaat, David and Masui, Toshihiko and Rogelj, Joeri and Strefler, Jessica and Drouet, Laurent and Krey, Volker and Luderer, Gunnar and Harmsen, Mathijs and Takahashi, Kiyoshi and Baumstark, Lavinia and Doelman, Jonathan C. and Kainuma, Mikiko and Klimont, Zbigniew and Marangoni, Giacomo and Lotze-Campen, Hermann and Obersteiner, Michael and Tabeau, Andrzej and Tavoni, Massimo},
	month = jan,
	year = {2017},
	pages = {153--168},
}

@techreport{international_energy_agency_electricity_2024,
	title = {Electricity {Market} {Report} 2024: {Analysis} and {Forecast} to 2026},
	url = {https://iea.blob.core.windows.net/assets/18f3ed24-4b26-4c83-a3d2-8a1be51c8cc8/Electricity2024-Analysisandforecastto2026.pdf},
	institution = {International Energy Agency},
	author = {{International Energy Agency}},
	year = {2024},
}

@article{buhler_comparative_2019,
	title = {A comparative assessment of electrification strategies for industrial sites: {Case} of milk powder production},
	volume = {250},
	issn = {03062619},
	shorttitle = {A comparative assessment of electrification strategies for industrial sites},
	doi = {10.1016/j.apenergy.2019.05.071},
	 
	urldate = {2024-12-04},
	journal = {Applied Energy},
	author = {Bühler, Fabian and Zühlsdorf, Benjamin and Nguyen, Tuong-Van and Elmegaard, Brian},
	month = sep,
	year = {2019},
	pages = {1383--1401},
}

@article{wiertzema_bottomup_2020,
	title = {Bottom–{Up} {Assessment} {Framework} for {Electrification} {Options} in {Energy}-{Intensive} {Process} {Industries}},
	volume = {8},
	issn = {2296-598X},
	doi = {10.3389/fenrg.2020.00192},
	urldate = {2024-12-04},
	journal = {Frontiers in Energy Research},
	author = {Wiertzema, Holger and Svensson, Elin and Harvey, Simon},
	month = aug,
	year = {2020},
	pages = {192},
}

@article{lopez_towards_2023,
	title = {Towards defossilised steel: {Supply} chain options for a green {European} steel industry},
	volume = {273},
	issn = {03605442},
	shorttitle = {Towards defossilised steel},
	doi = {10.1016/j.energy.2023.127236},
	 
	urldate = {2024-12-04},
	journal = {Energy},
	author = {Lopez, Gabriel and Galimova, Tansu and Fasihi, Mahdi and Bogdanov, Dmitrii and Breyer, Christian},
	month = jun,
	year = {2023},
	pages = {127236},
}

@article{dragomir_empirical_2023,
	title = {Empirical {Assessment} of {Carbon} {Reduction} and {Energy} {Transition} {Targets} of {European} {Companies}},
	volume = {17},
	copyright = {http://creativecommons.org/licenses/by/4.0},
	issn = {2558-9652},
	doi = {10.2478/picbe-2023-0067},
	number = {1},
	urldate = {2024-12-04},
	journal = {Proceedings of the International Conference on Business Excellence},
	author = {Dragomir, Voicu D. and Dumitru, Mădălina and Duţescu, Adriana and Perevoznic, Mădălina Florentina},
	month = jul,
	year = {2023},
	pages = {718--727},
}

@article{gerres_review_2019,
	title = {A review of cross-sector decarbonisation potentials in the {European} energy intensive industry},
	volume = {210},
	issn = {09596526},
	doi = {10.1016/j.jclepro.2018.11.036},
	 
	urldate = {2024-12-04},
	journal = {Journal of Cleaner Production},
	author = {Gerres, Timo and Chaves Ávila, José Pablo and Llamas, Pedro Linares and San Román, Tomás Gómez},
	month = feb,
	year = {2019},
	pages = {585--601},
}

@article{bataille_review_2018,
	title = {A review of technology and policy deep decarbonization pathway options for making energy-intensive industry production consistent with the {Paris} {Agreement}},
	volume = {187},
	issn = {09596526},
	doi = {10.1016/j.jclepro.2018.03.107},
	urldate = {2024-12-06},
	journal = {Journal of Cleaner Production},
	author = {Bataille, Chris and Åhman, Max and Neuhoff, Karsten and Nilsson, Lars J. and Fischedick, Manfred and Lechtenböhmer, Stefan and Solano-Rodriquez, Baltazar and Denis-Ryan, Amandine and Stiebert, Seton and Waisman, Henri and Sartor, Oliver and Rahbar, Shahrzad},
	month = jun,
	year = {2018},
	pages = {960--973},
}

@article{geels_technological_2002,
	title = {Technological transitions as evolutionary reconfiguration processes: a multi-level perspective and a case-study},
	volume = {31},
	copyright = {https://www.elsevier.com/tdm/userlicense/1.0/},
	issn = {00487333},
	shorttitle = {Technological transitions as evolutionary reconfiguration processes},
	doi = {10.1016/S0048-7333(02)00062-8},
	 
	number = {8-9},
	urldate = {2024-12-06},
	journal = {Research Policy},
	author = {Geels, Frank W.},
	month = dec,
	year = {2002},
	pages = {1257--1274},
}

@article{trianni_framework_2014,
	title = {A framework to characterize energy efficiency measures},
	volume = {118},
	issn = {03062619},
	doi = {10.1016/j.apenergy.2013.12.042},
	 
	urldate = {2024-12-06},
	journal = {Applied Energy},
	author = {Trianni, Andrea and Cagno, Enrico and De Donatis, Alessio},
	month = apr,
	year = {2014},
	pages = {207--220},
}

@article{luderer_impact_2021,
	title = {Impact of declining renewable energy costs on electrification in low-emission scenarios},
	volume = {7},
	issn = {2058-7546},
	doi = {10.1038/s41560-021-00937-z},
	 
	number = {1},
	urldate = {2024-12-17},
	journal = {Nature Energy},
	author = {Luderer, Gunnar and Madeddu, Silvia and Merfort, Leon and Ueckerdt, Falko and Pehl, Michaja and Pietzcker, Robert and Rottoli, Marianna and Schreyer, Felix and Bauer, Nico and Baumstark, Lavinia and Bertram, Christoph and Dirnaichner, Alois and Humpenöder, Florian and Levesque, Antoine and Popp, Alexander and Rodrigues, Renato and Strefler, Jessica and Kriegler, Elmar},
	month = nov,
	year = {2021},
	pages = {32--42},
}

@article{acemoglu_advanced_2023,
	title = {Advanced {Technology} {Adoption}: {Selection} or {Causal} {Effects}?},
	volume = {113},
	issn = {2574-0768, 2574-0776},
	shorttitle = {Advanced {Technology} {Adoption}},
	doi = {10.1257/pandp.20231037},
	urldate = {2024-12-17},
	journal = {AEA Papers and Proceedings},
	author = {Acemoglu, Daron and Anderson, Gary and Beede, David and Buffington, Catherine and Childress, Eric and Dinlersoz, Emin and Foster, Lucia and Goldschlag, Nathan and Haltiwanger, John and Kroff, Zachary and Restrepo, Pascual and Zolas, Nikolas},
	month = may,
	year = {2023},
	pages = {210--214},
}

@incollection{acemoglu_automation_2024,
	title = {Automation and the {Workforce}: {A} {Firm}-{Level} {View} from the 2019 {Annual} {Business} {Survey}},
	url = {http://www.nber.org/chapters/c14741},
	booktitle = {Technology, {Productivity}, and {Economic} {Growth}},
	publisher = {University of Chicago Press},
	author = {Acemoglu, Daron and Anderson, Gary W. and Beede, David N. and Buffington, Catherine and Childress, Eric E. and Dinlersoz, Emin and Foster, Lucia S. and Goldschlag, Nathan and Haltiwanger, John C. and Kroff, Zachary and Restrepo, Pascual and Zolas, Nikolas},
	month = may,
	year = {2024},
}

@article{williams_technology_2012,
	title = {The {Technology} {Path} to {Deep} {Greenhouse} {Gas} {Emissions} {Cuts} by 2050: {The} {Pivotal} {Role} of {Electricity}},
	volume = {335},
	copyright = {http://www.sciencemag.org/site/feature/contribinfo/prep/license.xhtml},
	issn = {0036-8075, 1095-9203},
	shorttitle = {The {Technology} {Path} to {Deep} {Greenhouse} {Gas} {Emissions} {Cuts} by 2050},
	doi = {10.1126/science.1208365},
	number = {6064},
	urldate = {2024-12-17},
	journal = {Science},
	author = {Williams, James H. and DeBenedictis, Andrew and Ghanadan, Rebecca and Mahone, Amber and Moore, Jack and Morrow, William R. and Price, Snuller and Torn, Margaret S.},
	month = jan,
	year = {2012},
	pages = {53--59},
}

@article{capros_energy-system_2019,
	title = {Energy-system modelling of the {EU} strategy towards climate-neutrality},
	volume = {134},
	issn = {03014215},
	doi = {10.1016/j.enpol.2019.110960},
	 
	urldate = {2024-12-17},
	journal = {Energy Policy},
	author = {Capros, Pantelis and Zazias, Georgios and Evangelopoulou, Stavroula and Kannavou, Maria and Fotiou, Theofano and Siskos, Pelopidas and De Vita, Alessia and Sakellaris, Konstantinos},
	month = nov,
	year = {2019},
	pages = {110960},
}

@article{rissman_technologies_2020,
	title = {Technologies and policies to decarbonize global industry: {Review} and assessment of mitigation drivers through 2070},
	volume = {266},
	issn = {03062619},
	shorttitle = {Technologies and policies to decarbonize global industry},
	doi = {10.1016/j.apenergy.2020.114848},
	 
	urldate = {2024-12-17},
	journal = {Applied Energy},
	author = {Rissman, Jeffrey and Bataille, Chris and Masanet, Eric and Aden, Nate and Morrow, William R. and Zhou, Nan and Elliott, Neal and Dell, Rebecca and Heeren, Niko and Huckestein, Brigitta and Cresko, Joe and Miller, Sabbie A. and Roy, Joyashree and Fennell, Paul and Cremmins, Betty and Koch Blank, Thomas and Hone, David and Williams, Ellen D. and De La Rue Du Can, Stephane and Sisson, Bill and Williams, Mike and Katzenberger, John and Burtraw, Dallas and Sethi, Girish and Ping, He and Danielson, David and Lu, Hongyou and Lorber, Tom and Dinkel, Jens and Helseth, Jonas},
	month = may,
	year = {2020},
	pages = {114848},
}

@article{gailani_assessing_2024,
	title = {Assessing the potential of decarbonization options for industrial sectors},
	volume = {8},
	issn = {25424351},
	doi = {10.1016/j.joule.2024.01.007},
	 
	number = {3},
	urldate = {2024-12-17},
	journal = {Joule},
	author = {Gailani, Ahmed and Cooper, Sam and Allen, Stephen and Pimm, Andrew and Taylor, Peter and Gross, Robert},
	month = mar,
	year = {2024},
	pages = {576--603},
}

@article{madeddu_co2_2020,
	title = {The {CO}$_{\textrm{2}}$ reduction potential for the {European} industry via direct electrification of heat supply (power-to-heat)},
	volume = {15},
	issn = {1748-9326},
	doi = {10.1088/1748-9326/abbd02},
	number = {12},
	urldate = {2024-12-17},
	journal = {Environmental Research Letters},
	author = {Madeddu, Silvia and Ueckerdt, Falko and Pehl, Michaja and Peterseim, Juergen and Lord, Michael and Kumar, Karthik Ajith and Krüger, Christoph and Luderer, Gunnar},
	month = dec,
	year = {2020},
	pages = {124004},
}

@misc{IEA2022,
  author       = {{IEA}},
  year         = {2022},
  title        = {Hungary 2022},
  institution  = {International Energy Agency (IEA)},
  address      = {Paris},
  url          = {https://www.iea.org/reports/hungary-2022},
  note         = {Accessed January 28, 2025},
}

@misc{reisch_rewiring,
  author       = {Tobias Reisch and András Borsos and Stefan Thurner},
  title        = {Supply chain network rewiring dynamics at the firm-level},
  year         = {2025},
  eprint       = {2503.20594},
  archivePrefix= {arXiv},
  primaryClass = {econ.GN},
  doi          = {10.48550/arXiv.2503.20594},
  url          = {https://arxiv.org/abs/2503.20594}
}

@misc{statistik_austria_systematik_2008,
  title        = {Systematik der {Wirtschaftstätigkeiten} Ö{NACE} 2008},
  url          = {https://www.statistik.at/fileadmin/publications/Systematik_der_Wirtschaftstaetigkeiten__OENACE_2008.pdf},
  publisher    = {Statistik Austria},
  author       = {{Statistik Austria}},
  note         = {Accessed January 28, 2025},
  year         = {2008},
}

@article{Huber1964,
  title = {Robust Estimation of a Location Parameter},
  volume = {35},
  ISSN = {0003-4851},
  url = {http://dx.doi.org/10.1214/aoms/1177703732},
  DOI = {10.1214/aoms/1177703732},
  number = {1},
  journal = {The Annals of Mathematical Statistics},
  publisher = {Institute of Mathematical Statistics},
  author = {Huber,  Peter J.},
  year = {1964},
  month = mar,
  pages = {73–101}
}

@article{seabold_statsmodels_2010,
	title = {Statsmodels: {Econometric} and {Statistical} {Modeling} with {Python}},
	shorttitle = {Statsmodels},
	doi = {10.25080/Majora-92bf1922-011},
	urldate = {2025-09-14},
	journal = {scipy},
	author = {Seabold, Skipper and Perktold, Josef},
	month = may,
	year = {2010},
}

\section*{Author Contributions}
J.S. and S.T. conceived the work. J.S. and A.B. cleaned and prepared the data. J.S. and A.B. wrote the code. J.S., A.B., S.T. performed the data analysis. All authors analyzed and interpreted the results. J.S. and S.T. wrote the paper. All authors contributed towards the final manuscript.

\section*{Competing Interests}
The authors declare no competing interests.

\section*{Acknowledgments}
We are grateful to Jan Hurt for insightful discussions and valuable feedback provided throughout various stages of this study.

\newpage


\onecolumngrid

\pagebreak[4]

\section*{Supplementary Information}
\section*{Supplementary Methods}
\subsection*{Electricity mix of Hungary and forecast until 2050}
\label{SI:electricity_mix_forecast}

\begin{figure*}[htb]
\centering
\includegraphics[width=12.0cm]{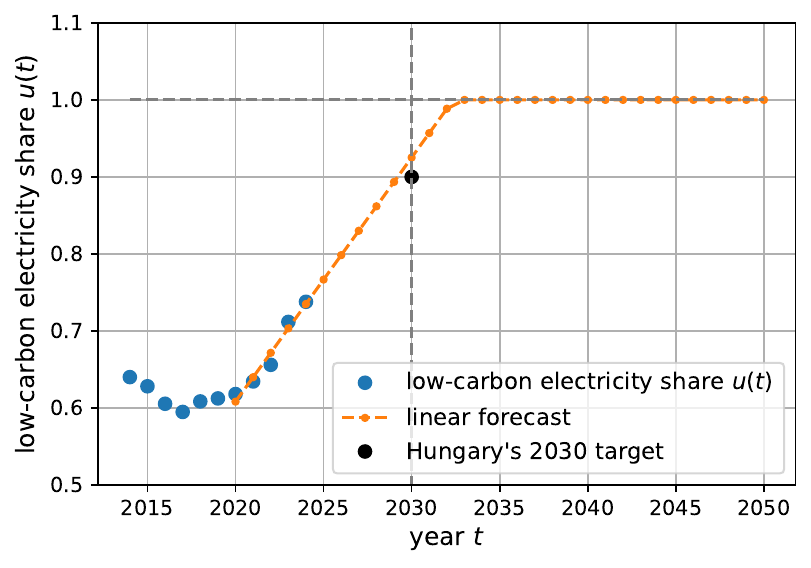}
\caption{Low-carbon share of the electricity mix of Hungary $u(t)$ for the years 2014-2024 and a forecast for the evolution of the low-carbon share until 2050 $u_{\text{forecast}}(t)$ based on a linear regression for the years 2020-2024. Hungary has set a target of 90\% low-carbon electricity generation by 2030 \cite{ministry_for_innovation_and_technology_national_2020}.}
\label{SI-FIG.:electricity_mix}
\end{figure*}

We use data from the online platform Ember, which provides annual data on clean and fossil electricity generation in terawatt-hours, to calculate the low-carbon share of Hungary's annual electricity mix $u(t)$ \cite{ember_electricity_2023}. To estimate the future low-carbon share, we perform a linear regression based on the 2020-2024 observation period, during which the low-carbon share increased by 10\%. This scenario, which would enable Hungary to reach its self-proclaimed target of 90\% low-carbon electricity generation by 2030, requires an annual increase of $u(t)$ by 3.2\%. In this scenario, Hungary’s electricity grid would be essentially decarbonized by the year 2033. Supplementary Tab. \ref{SI-Tab.:low-carbon_share_Hungary} and Supplementary Fig.\ref{SI-FIG.:electricity_mix} provide an overview of the evolution and the forecast of the low-carbon share $u(t)$ based on the electricity generation data.

\subsection*{Electricity, gas and fuel price evolution in Hungary}
\label{SI:electricity_gas_prices}
\begin{figure*}[htb]
\centering
\includegraphics[width=12.0cm]{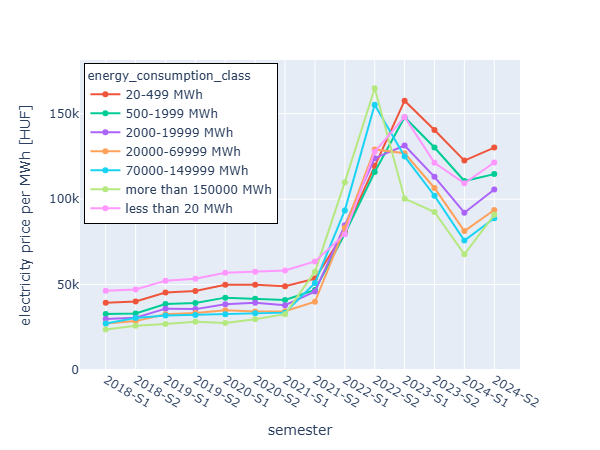}
\caption{Semi-annual electricity prices for non-household consumers in Hungary from 2018 to 2024, categorized by size classes of energy consumers. Size classes are defined based on annual electricity consumption, divided into the following seven consumption bands: less than 20 MWh, 20–499 MWh, 500–1,999 MWh, 2,000–19,999 MWh, 20,000–69,999 MWh, 70,000–149,999 MWh, and more than 150,000 MWh. Data is sourced from EUROSTAT.}
\label{SI-FIG.:electricity_price}
\end{figure*}

\begin{figure*}[htb]
\centering
\includegraphics[width=12.0cm]{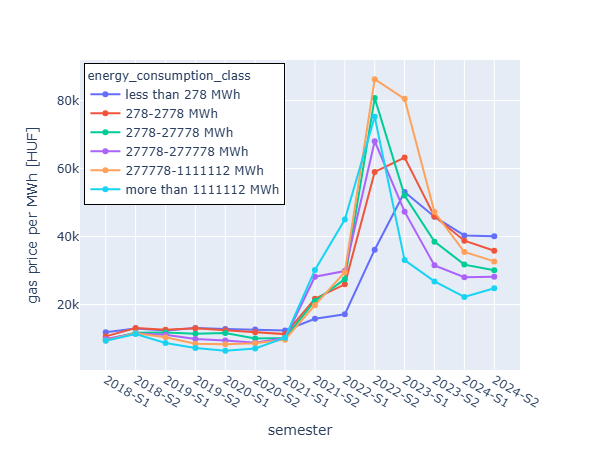}
\caption{Semi-annual gas prices for non-household consumers in Hungary from 2018 to 2024, categorized by size classes of energy consumers. Size classes are defined based on annual gas consumption, divided into the following sixconsumption bands: less than 278 MWh, 278–2,778 MWh, 2,778–27,778 MWh, 27,778–277,778 MWh, 277,778–1,111,112 MWh, and more than 1,111,112 MWh. The size classes are converted from Gigajoules in the original dataset to megawatt-hours. Data is sourced from EUROSTAT \cite{eurostat_gas_2024}.}
\label{SI-FIG.:gas_price}
\end{figure*}

Electricity and gas prices for non-household consumers are sourced from EUROSTAT \cite{eurostat_electricity_2024}\cite{eurostat_gas_2024}. These data, based on the reports of energy providers, reflect price trends across various energy consumption classes. Electricity prices are categorized into seven ranges, while gas prices are divided into six. The prices, provided in Hungarian Forint (HUF), are reported semi-annually. Supplementary Figs. \ref{SI-FIG.:electricity_price} and \ref{SI-FIG.:gas_price} depict the evolution of the electricity and the gas price for non-household consumers in Hungary. Since the supply chain network data are aggregated on a semi-annual basis, we can directly use the semi-annual price data to convert monetary values into kilowatt-hours of energy consumed. For firms with annual reporting requirements, however, we apply the average of the semi-annual prices within the year. This procedure may still result in some under- or overestimation. \\
Since the price data reflect only averages per energy consumption range, our estimates may be imprecise for firms whose actual prices differ due to their specific sourcing strategies, such as retail purchases, power purchase agreements, or spot market trading. \\

\begin{figure*}[htb]
\centering
\includegraphics[width=16.0cm]{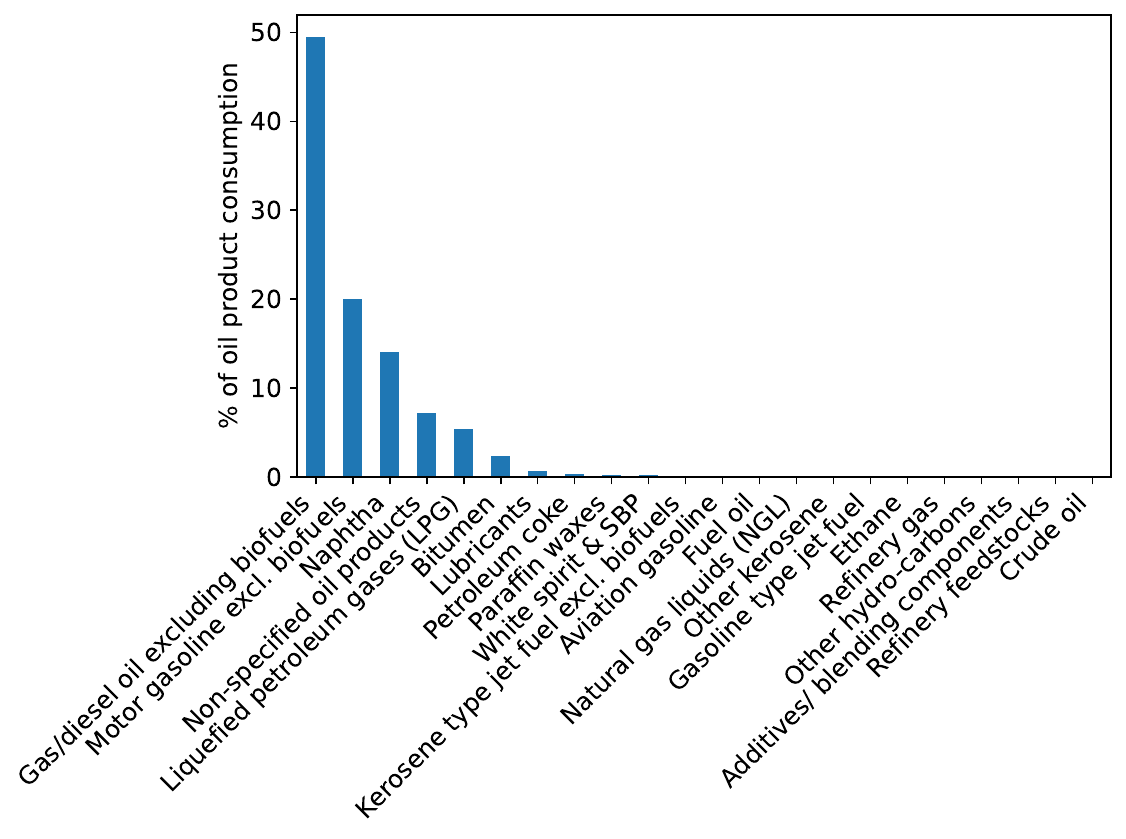}
\caption{Percantages of annual consumption of oil products for the year 2023 as derived from final consumption in the National Detailed Energy Balance for Hungary \cite{mekh_energy_balance_2025}}
\label{SI-FIG.:oil_consumption}
\end{figure*}

{In order to estimate oil consumption,} we assume that firms primarily consume oil in the form of fuels, using fuel price trends as a proxy to convert observed oil expenditures into energy units. Data from the National Detailed Energy Balance, provided by the Hungarian Energy and Public Utility Regulatory Authority, confirms that diesel and gasoline are by far the most significant forms of oil product consumption \cite{mekh_energy_balance_2025}. Supplementary Fig. \ref{SI-FIG.:oil_consumption} illustrates the distribution of oil product consumption for 2023, derived from the National Detailed Energy Balance data. Diesel accounts for nearly half of oil product consumption, followed by gasoline at 20\%. Naphtha represents 14\% of consumption; however, according to the National Detailed Energy Balance, the chemical and petrochemical industries are the sole consumers of naphtha.

\begin{figure*}[htb]
\centering
\includegraphics[width=12.0cm]{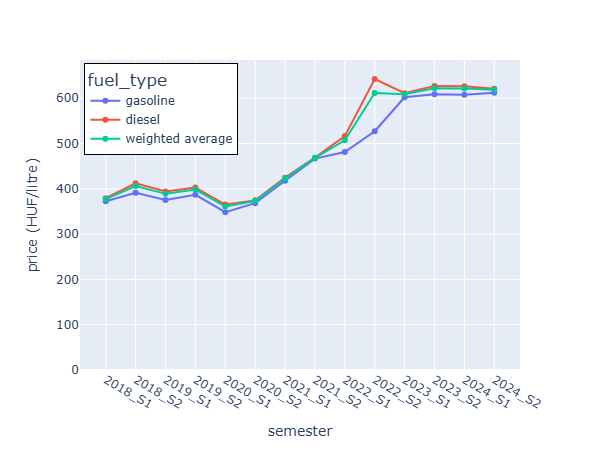}
\caption{ Semi-annual fuel prices for non-household consumers in Hungary between 2018-2024. Prices for gasoline, diesel, and a weighted average of the two are presented. Data is sourced from \cite{eu_oil_bulletin}.}
\label{SI-FIG.:fuel_price}
\end{figure*}

Data on fuel prices is obtained from the historical price trends in the Weekly Oil Bulletin provided by the EU \cite{eu_oil_bulletin}. To determine a unique price for oil products, we calculate a weighted average of gasoline and diesel prices in Hungary. The weights are based on the relative consumption of gasoline and diesel in Hungary, also derived from the Weekly Oil Bulletin data, implicitly assuming that firms consume these fuels in similar proportions. In general, diesel represents about 74\% of fuel consumption, whereas gasoline represents about 26\% during the observation period from 2020 to 2024. We arrive at a unified price for fuel consumption in Hungary that allows us to convert monetary inputs for oil products into kilowatt-hours consumed. Supplementary Fig. \ref{SI-FIG.:fuel_price} depicts the evolution of fuel prices for gasoline, diesel, and their weighted average in Hungary for each semester from 2018 to 2024. Fuel prices have increased significantly since 2021, reflecting the broader energy crisis in Europe.

\subsection*{OLS vs. robust estimation}
\begin{figure*}[htb]
\centering
\includegraphics[width=13.0cm]{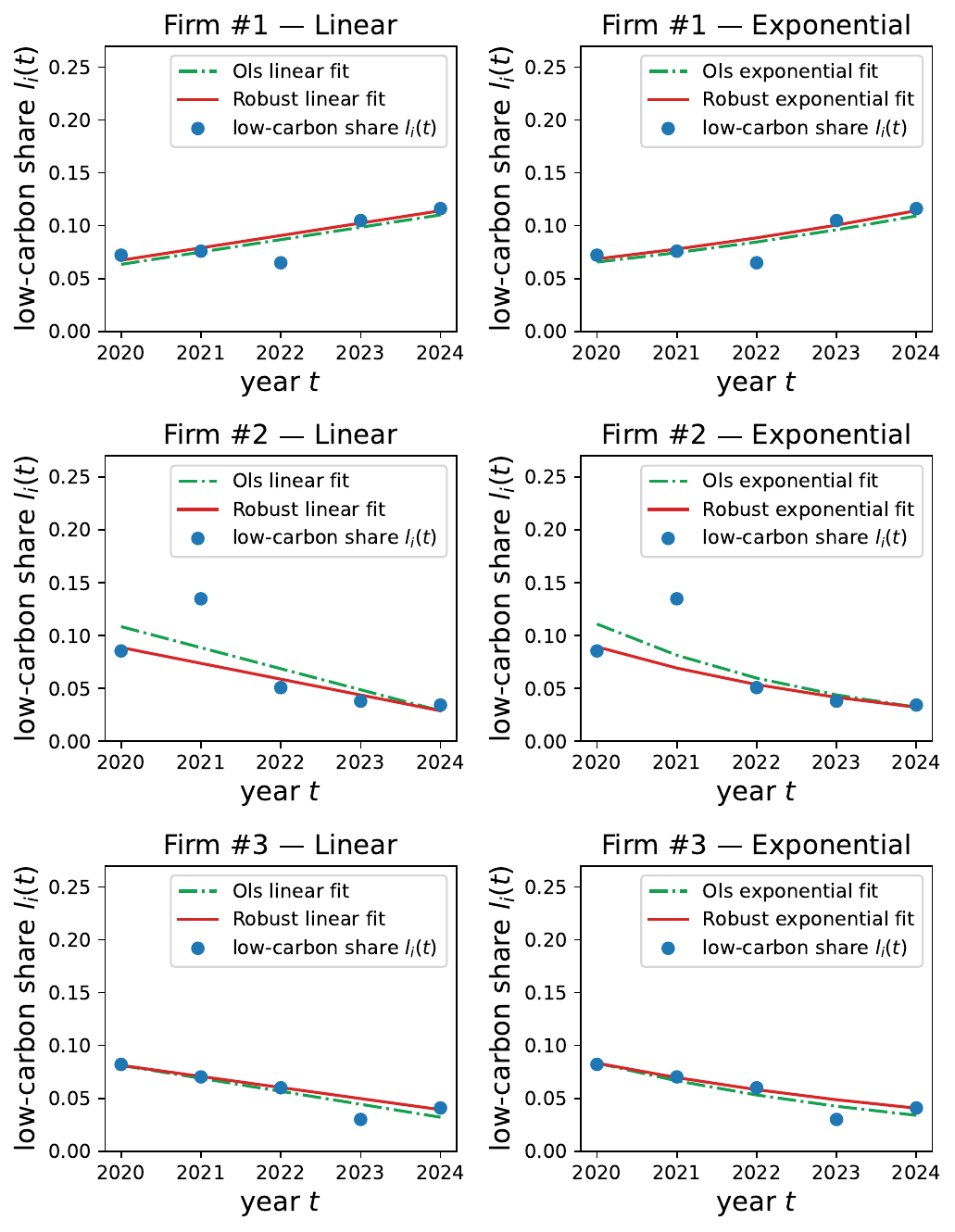}
\caption{Comparison of OLS and robust regression fits (Huber loss) for linear and exponential models of the low-carbon share, $l_i(t)$ across three example firms. The robust estimator reduces sensitivity to outliers (e.g. Firm \#2), while providing trends consistent with OLS when data are rather well-behaved (e.g. Firm Firm \#1 and Firm \#3)}
\label{SI-FIG.:ols_vs_robust}
\end{figure*}

Instead of an ordinary least squares (OLS) estimator, we employ a robust estimator based on the Huber norm \cite{Huber1964}, which down-weights outliers to obtain more reliable estimates of trends in the low-carbon share, $l_i(t)$. This approach is particularly valuable in the presence of crisis years within our observation period (2020–2024), such as the COVID-19 pandemic in 2020 or the energy crisis in 2022, and helps to mitigate potential firm-specific fluctuations. The details of the robust estimation procedure are provided in the Methods section of the main text. Supplementary Fig. \ref{SI-FIG.:ols_vs_robust} compares the decarbonization trends, $\delta_i$, and decarbonization rates, $\lambda_i$, estimated using OLS and robust regression. The robust estimator reduces sensitivity to outliers (e.g., Firm \#2) while yielding results that are consistent with OLS when the data are well-behaved (e.g., Firm \#1 and Firm \#3).

\section*{Supplementary Discussion}
\subsection*{Firm sample description}
\label{SI:firm_sample_description}
We compare our constructed firm sample of 25,231 firms with the total firm dataset, which we treat as the 'ground truth,' to evaluate how well the sample represents the overall firm population in terms of energy inputs and revenue. Specifically, we compare the aggregated monetary inputs for gas, electricity, and oil, as well as the total revenue of firms in our sample, with the corresponding aggregated values in the total dataset. To ensure a fair comparison, we exclude energy-providing sectors, firms from the financial sector, firms covered by the emission trading system (ETS) and firms with no NACE category from the 'ground truth' dataset. However, we do not apply the time series consistency restrictions for energy inputs and revenue outlined in the Methods section on Firm sample construction. As a result, we exclude firms in the following NACE 4-digit categories from the 'ground truth': 'D35.1 - Electric power generation, transmission, and distribution,' 'D35.1.1 - Production of electricity,' 'D35.1.2 - Transmission of electricity,' 'D35.1.3 - Distribution of electricity,' 'D35.1.4 - Trade of electricity,' 'D35.2.1 - Manufacture of gas,' 'D35.2.2 - Distribution of gaseous fuels through mains,' 'D35.2.3 - Trade of gas through mains,' 'B6.1.0 - Extraction of crude petroleum,' 'C19.2.0 - Manufacture of refined petroleum products,' 'G47.3.0 - Retail sale of automotive fuel in specialized stores,' 'G46.7.1 - Wholesale of solid, liquid, and gaseous fuels and related products,', 'K - Financial and insurance activities,' and 'H52.2.1 - Service activities incidental to land transportation'.

\begin{figure*}[htb]
\centering
\includegraphics[width=10.0cm]{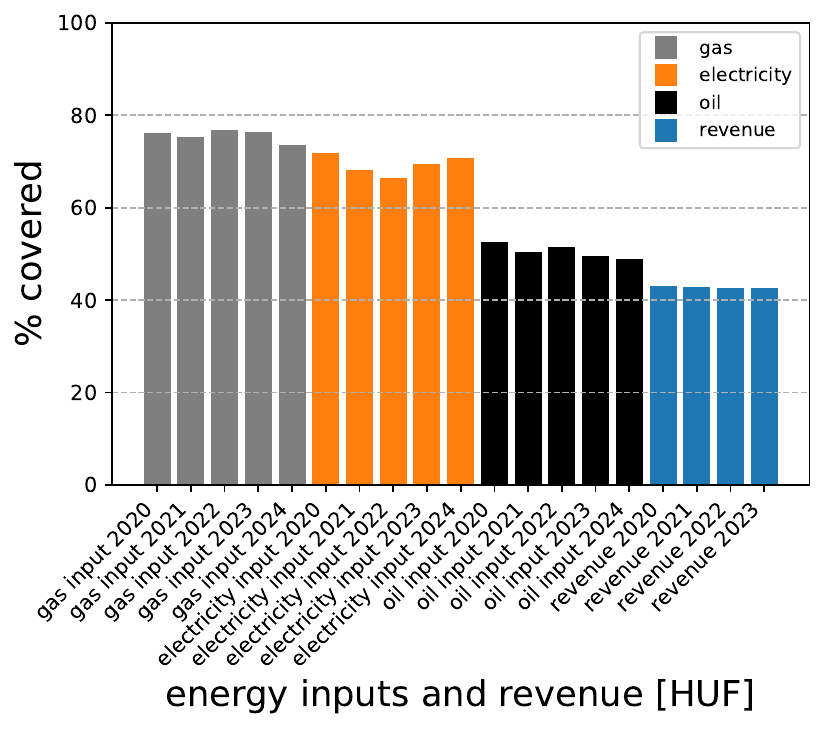}
\caption{Bar plot showing the percentage coverage of the constructed firm sample relative to the total firm dataset from 2020 to 2023. The chart presents the share of aggregated gas, electricity, and oil inputs (in HUF), as well as total revenue, captured by the constructed sample compared to the overall dataset.}
\label{SI-FIG.:coverage_sample}
\end{figure*}

This results in a 'ground truth' dataset of 434,988 firms, which includes all firms regardless the continuity of energy inputs or revenue, except those in the excluded NACE categories or firms covered in the ETS. We then calculate the covered energy shares and revenue as fractions, defined as the ratio between the sums of the respective variables in our reconstructed firm sample and those in the 'ground truth' dataset. Supplementary Fig. \ref{SI-FIG.:coverage_sample} illustrates the coverage, and Supplementary Tab. \ref{SI-Tab.:coverage_energy_inputs_revenue} provides further details. We achieve very good coverage for gas and electricity, with 75\% coverage for gas inputs and approximately 70\% for electricity inputs across all years. We also maintain good coverage for oil and revenue, with approximately 50\% coverage for oil inputs across 2020-2024 and approximately 43\% coverage for total revenue across 2020-2023. This means that our constructed firm sample captures a significant share of energy inputs and revenue from the total firm dataset, indicating that the sample consists of large and thus relevant firms.

To provide a more detailed description of our sample, we present the aggregated energy purchases (gas, oil, and electricity) of all firms in the sample from 2020 to 2024, along with their conversion into energy units (terawatt-hours, TWh) using energy prices, as shown in Supplementary Fig. \ref{SI-FIG.:energy_purchases}. As illustrated in Supplementary Fig. \ref{SI-FIG.:energy_purchases}a, energy purchases increased significantly in 2022 and 2023, coinciding with peak prices for gas, oil, and electricity. However, when these purchases are converted into energy units using energy prices, it becomes clear that total energy consumption across the different energy types remained relatively stable for the firms in the sample,  as shown in Supplementary Fig. \ref{SI-FIG.:energy_purchases}b. Notably, electricity consumption appears to have declined since 2022. 

\subsection*{Comparison of firm sample to sectoral energy consumption data}
\label{SI:comparison_firm_sample}

We assess how well our firm sample represents sectoral energy consumption in Hungary by comparing it to official statistics. Specifically, we examine the total energy consumption (in TWh) of gas, electricity, and oil for the NACE 1-digit sectors 'A - Agriculture, forestry and fishing,' 'B - Mining and quarrying,' 'C - Manufacturing,' and 'F - Construction' using data from Hungary’s energy balance \cite{mekh_energy_balance_2025}. Supplementary Fig. \ref{SI-FIG.:energy_consumption_comparison} presents time series data for both the official statistics and our estimated energy consumption for the aggregated firm sample. Since our sample consists of only 25,231 firms, it is not expected to cover the total energy consumption of each sector. However, it should capture the relative consumption levels across sectors to some degree. As shown in Supplementary Fig.\ref{SI-FIG.:energy_consumption_comparison}, the sectoral distribution of gas, electricity, and oil consumption is well reflected in our estimates, with 'C - Manufacturing' consuming the most energy, followed by 'A - Agriculture, forestry and fishing,' and 'B - Mining and quarrying' consuming the least. However, our estimates indicate higher gas consumption and lower electricity consumption for 'C - Manufacturing' compared to official statistics. This suggests a possible underestimation of electricity consumption and an overestimation of gas consumption in our firm sample. 

To further evaluate the accuracy of our estimates, we compare the share of electricity consumption in each NACE 1-digit sector using official energy balance data \cite{mekh_energy_balance_2025}. For each year between 2020 and 2024, we calculate the electricity share in the sectoral energy mix by dividing electricity consumption by the total consumption of electricity, natural gas, and oil. We then compare these shares to those in our firm sample. As shown in Supplementary Fig.\ref{SI-FIG.:electricity_shares_comparison}, our estimates generally capture the relative importance of electricity across sectors. 'C - Manufacturing' has the highest share of electricity in the energy mix, consistent with official statistics. However, our estimates show a lower overall electricity share, indicating a potential underestimation of electricity consumption. The discrepancy is particularly pronounced in 2022.

We additionally evaluate the same statistics for NACE 2-digit subsectors of the manufacturing sector, for which official data are available from Hungary’s final energy use in industry for 2020–2023 \cite{mekh_industry_final_energy_2025}. The sector labels in the official statistics were manually matched to NACE 2-digit codes. While the correspondence is largely one-to-one, some sectors represent aggregates of several NACE 2-digit codes. Supplementary Tab. \ref{SI-Tab.:manufacturing_nace} documents these correspondences. The relative importance of oil, gas, and electricity consumption across most manufacturing subsectors is preserved in our sample (Supplementary Fig.~\ref{SI-FIG.:manufacuring_energy_consumption_comparison}). Gas consumption is reasonably well reflected, whereas electricity appears underrepresented. As expected, the chemical sector is not well covered in oil use, since it also consumes oil in the form of naphtha and other products not captured in our analysis. Supplementary Fig. \ref{SI-FIG.:manufacuring_electricity_shares_comparison} further compares electricity shares of manufacturing NACE 2-digit subsectors between the official statistics and our firm sample. While some of the relative importance across sectors is retained, the under-representation of electricity has a sizable impact on aggregate electricity shares. We suspect that firms in this sector may also procure electricity from the spot market, which cannot be captured by our method of assigning energy providers based on their NACE codes. Additionally, our analysis relies on a restricted sample of firms, and it remains inherently uncertain how much of the gap in electricity use relative to official statistics can be attributed to firms not captured in our data

\subsection*{Influence of suppliers and customers on transitioning behaviour}
We investigate firms’ transition behavior, and how it is associated with the transition behavior of their suppliers and customers. Our analysis consists of two parts.  

First, we compute the mean decarbonization trend of each firm’s suppliers and customers, based on their individual trends $\overline{\delta_j}$, and correlate this mean with the firm’s own decarbonization trend, $\delta_i$. A positive correlation indicates that the decarbonization trends of firms and their partners are aligned (both increasing or both decreasing), whereas a negative correlation indicates that firms and their partners tend to move in opposite directions. We report Pearson and Spearman correlations in Supplementary Fig.~\ref{SI-FIG.:transition_cascade_trend}. The results show negative correlations between the mean trends of suppliers and focal firms across tiers, positive correlations between mean trends of customers and focal firms for tiers 1 and 2, and negative correlations for customers at tier 3. While the correlations are not large, the pattern suggests that firms’ decarbonization trends are more closely associated with those of their direct customers than with those of their suppliers.
 
Second, we calculate the share of suppliers and customers that are transitioning for each firm, and correlate this share with the firm’s own transition status. A positive correlation indicates that transitioning firms are associated with having a higher fraction of transitioning partners than non-transitioning firms, while a negative correlation indicates the opposite. We report Pearson and Spearman correlations in Fig.~\ref{SI-FIG.:transition_cascade_share}. The results again show negative correlations between the share of transitioning suppliers and firms’ own transition status, and positive correlations between the share of transitioning direct customers and firms’ own transition status. This pattern suggests that firms’ transition decisions are more closely associated with the behavior of their customers than with that of their suppliers, which is consistent with the interpretation that customer demand creates pressure to decarbonize.

These findings should be interpreted with caution. Because we only observe the transition status of firms in our sample, our measures of partner behavior are necessarily incomplete and exclude suppliers and customers outside the dataset. Moreover, the number of partners increases sharply with distance in the network, from typically around 100 in tier 1 to several thousand in tier 2 and more than 10,000 in tier 3. This scaling may influence the correlations and complicates comparisons across tiers. We therefore regard this analysis as exploratory, and it serves as motivation for future research on the relationship between firms’ participation in the energy transition and the behavior of their suppliers and customers.

\subsection*{Uncertainty analysis of the energy scenarios}

To assess the uncertainty of the energy scenarios, we apply a leave-one-year-out procedure, analogous to bootstrapping, in which decarbonization trends, $\delta_i$, and decarbonization rates, $\lambda_i$, are re-estimated after sequentially omitting one year from the sample. For each energy scenario—business-as-usual (linear), business-as-usual (exponential), transition (linear), and transition (exponential)—this results in five runs: one main run using all years 2020–2024 and four additional runs, each excluding a different year. This approach allows us to construct an uncertainty envelope around the main run, defined by the minimum and maximum values of the low-carbon share, $l_i$, across all runs. Supplementary Fig. \ref{SI-FIG.:uncertainty_scenarios} illustrates the four energy scenarios, the leave-one-year-out runs, and the resulting uncertainty envelopes. Supplementary Tab. \ref{tab:scenarios_uncertainty} reports the values of the main runs alongside the lower and upper bounds across all leave-one-year-out runs. Overall, the scenarios are remarkably stable under the leave-one-year-out analysis. The linear scenarios show only small differences between their minimum and maximum values. The exponential scenarios are also stable, though their upper bounds deviate more strongly from the main runs. This deviation arises when the year 2020 is excluded from the estimation, which increases the slope of some firms and leads to an uptick in the projected low-carbon share. This is intuitive, as the estimated low-carbon share in 2020 is higher than in the following years (2020–2024 observed period), so including 2020 in the estimation leads to more negative slopes. Despite this sensitivity, the close agreement of aggregate low-carbon shares across the different runs underscores the robustness and stability of our results.

Although we test the sensitivity of our results to the estimation period, we do not explicitly examine the assumption of extended average energy consumption, as introducing additional assumptions would substantially increase the complexity of interpreting the results. The scenario runs are intended primarily for illustration: they show that current firm-level decarbonization trends are insufficient to meaningfully change energy consumption, whereas meaningful reductions would be achievable if firms followed the frontrunners within their industries.

\subsection*{Limitations in establishing causality between firm characteristics and transition behavior}

A natural question arising from our analysis is whether specific firm characteristics \emph{cause} firms to transition toward higher shares of low-carbon energy, or conversely, to remain locked-in to fossil-fuel–intensive portfolios. While our results identify patterns consistent with a lock-in mechanism, the empirical design does not allow us to determine the direction of causality.

\subsubsection*{Panel Granger causality tests}

To explore temporal precedence, we attempted to implement the heterogeneous panel Granger non-causality test proposed by Dumitrescu and Hurlin (2012)\cite{Dumitrescu2012}. In principle, this method can identify whether changes in firm characteristics systematically precede changes in low-carbon shares across firms.

In practice, however, the test requires a considerably longer time-series dimension, typically $T \geq 10$, for stable estimation. Our dataset contains only five yearly observations per firm, leaving too few effective data points once lags are included. As a result, many firm-level regressions become numerically unstable, causing the test statistic $Z_{\text{bar}}$ to diverge to unrealistically large values. In such cases, the p-values no longer provide meaningful evidence of causality but instead reflect the breakdown of the test under extremely short panels. Traditional time-series Granger tests suffer from the same limitation \cite{Granger1969}. A future study with a longer time series could meaningfully revisit this analysis.

\subsubsection*{Implications and future directions}

Given these limitations, our analysis should be viewed as descriptive: we document systematic differences between firms that increase their low-carbon energy share and those that do not, but we do not identify the mechanisms driving these differences. Future work could pursue causal identification by leveraging longer time-series data, natural or quasi-experiments (e.g., policy shocks, regulatory changes, eligibility thresholds), or event-study designs to examine dynamic responses to exogenous changes. Such approaches would provide a stronger foundation for causal claims about the determinants of firms’ energy transition pathways.

\subsection*{Estimation of the effect of solar PV self-consumption}
Our firm-level electricity consumption estimates are derived by aggregating monetary expenditures on electricity purchases from energy providers and converting them into kilowatt-hours. This measurement strategy does not capture behind-the-meter (BTM) electricity consumption, i.e.\ electricity generated and consumed on-site, which has become increasingly relevant due to the rapid expansion of solar photovoltaic (PV) capacity in Hungary during the observation period. PV installations on commercial premises commonly serve both self-consumption and grid feed-in.

To approximate the electricity we may be missing by ignoring BTM self-consumption, we draw on several complementary data sources.

The Hungarian Energy and Utilities Regulatory Authority (MEKH) publishes annual data of total gross PV electricity generation, denoted $G^{\mathrm{MEKH}}(t)$, which include both grid-injected electricity and estimated production from smaller systems \cite{MEKH_1_2025, MEKH_Methodology_2025}. In contrast, the European Network of Transmission System Operators for Electricity (ENTSO-E) reports only electricity that actually enters the grid, denoted $G^{\mathrm{grid}}(t)$, through its transparency platform \cite{ENTSOE2025}. Annual country-level aggregations of this dataset are provided by Energy Charts \cite{EnergyCharts2025}.

The Hungarian transmission system operator MAVIR provides a detailed breakdown of installed PV capacity \cite{MAVIR2025}, distinguishing:  
(i) household-scale systems (HMKE);  
(ii) utility-scale PV plants primarily feeding into the grid; and  
(iii) industrial-scale installations intended for self-consumption (SCTE).  
The same source reports annual PV capacity factors, ranging around 17\%. Based on this classification, we assume that utility-scale PV systems feed all generated electricity into the grid, that SCTE installations generate exclusively for self-consumption by firms, and that a fraction of HMKE systems located on commercial buildings also produce electricity used by firms directly.

Let $C^{\mathrm{SCTE}}(t)$ denote the installed annual SCTE capacity and $CF(t)$ the annual PV capacity factor. SCTE generation is then 
\[
G^{\mathrm{SCTE}}(t) = C^{\mathrm{SCTE}}(t)\cdot CF(t) \cdot 8760.
\]
Because electricity produced with SCTE systems is almost certainly consumed by industrial users, the resulting estimates of firms’ low-carbon electricity shares may be distorted for industrial firms that have invested in this type of on-site generation. 

A study by the market research company Forsense \cite{Forsense2024}, drawing on MAVIR data, provides a subdivision of total HMKE capacity into residential, $C^{\mathrm{HMKE,res}}(t)$, and total HMKE. From this we infer commercial HMKE capacity, $C^{\mathrm{HMKE,com}}(t)$. We adopt the conservative assumption that the entirety of commercial HMKE generation is self-consumed rather than fed into the grid. While plausible for larger industrial firms, this is unlikely to hold for smaller service-sector firms that typically both self-consume and feed into the grid. We use this restrictive assumption deliberately to derive an upper bound on the average measurement error arising from unobserved self-consumption. Commercial HMKE generation is thus defined as: 
\[
G^{\mathrm{HMKE,com}}(t) = C^{\mathrm{HMKE,com}}(t)\cdot CF(t) \cdot 8760.
\]
Total commercial PV self-consumption is therefore
\[
G^{\mathrm{SC,com}}(t) = G^{\mathrm{SCTE}}(t) + G^{\mathrm{HMKE,com}}(t),
\]
and its share in total national PV generation is
\[
s^{\mathrm{PV,com}}(t) = \frac{G^{\mathrm{SC,com}}(t)}{G^{\mathrm{MEKH}}(t)}.
\]
On average, this share amounts to roughly 27\% over the observation period.

To relate commercially self-consumed PV electricity to total commercial electricity use, let $F^{\mathrm{com}}(t)$ denote commercial electricity demand from the detailed national energy balance \cite{mekh_energy_balance_2025}. We compute this by subtracting residential consumption, $F^{\mathrm{res}}(t)$ from total final electricity consumption, $F(t)$:
\[
F^{\mathrm{com}}(t) = F(t) - F^{\mathrm{res}}(t).
\]

The share of commercial total electricity consumption met through self-consumed PV is then
\[
s^{\mathrm{com}}(t) = \frac{G^{\mathrm{SC,com}}(t)}{F^{\mathrm{com}}(t)},
\]
which rises from approximately 2.3\% in 2020 to 8.8\% in 2024. Table~\ref{tab:self-consumption-1} and Tab.~\ref{tab:self-consumption-2} summarize the underlying data and calculated results.

Although this increase is substantial, its absolute magnitude remains modest relative to total commercial electricity demand, and the average impact is therefore likely to be limited. The impact is likely heterogeneous across sectors: service-sector firms may exhibit comparatively higher self-consumption shares relative to their overall electricity use, whereas large industrial firms typically consume much larger volumes of electricity, making the contribution of PV self-consumption relatively small. Consequently, our estimates for large industrial electricity consumers are unlikely to be materially affected. This may not hold for firms that have invested in SCTE solar PV systems; however, because we cannot identify these firms, this remains a limitation of our analysis. By contrast, the low-carbon electricity shares, $l_i(t)$ may be understated for smaller service-sector firms and should therefore be interpreted with caution.

\FloatBarrier

\section*{Supplementary Figures}

\begin{figure*}[htb]
\centering
\includegraphics[width=16.0cm]{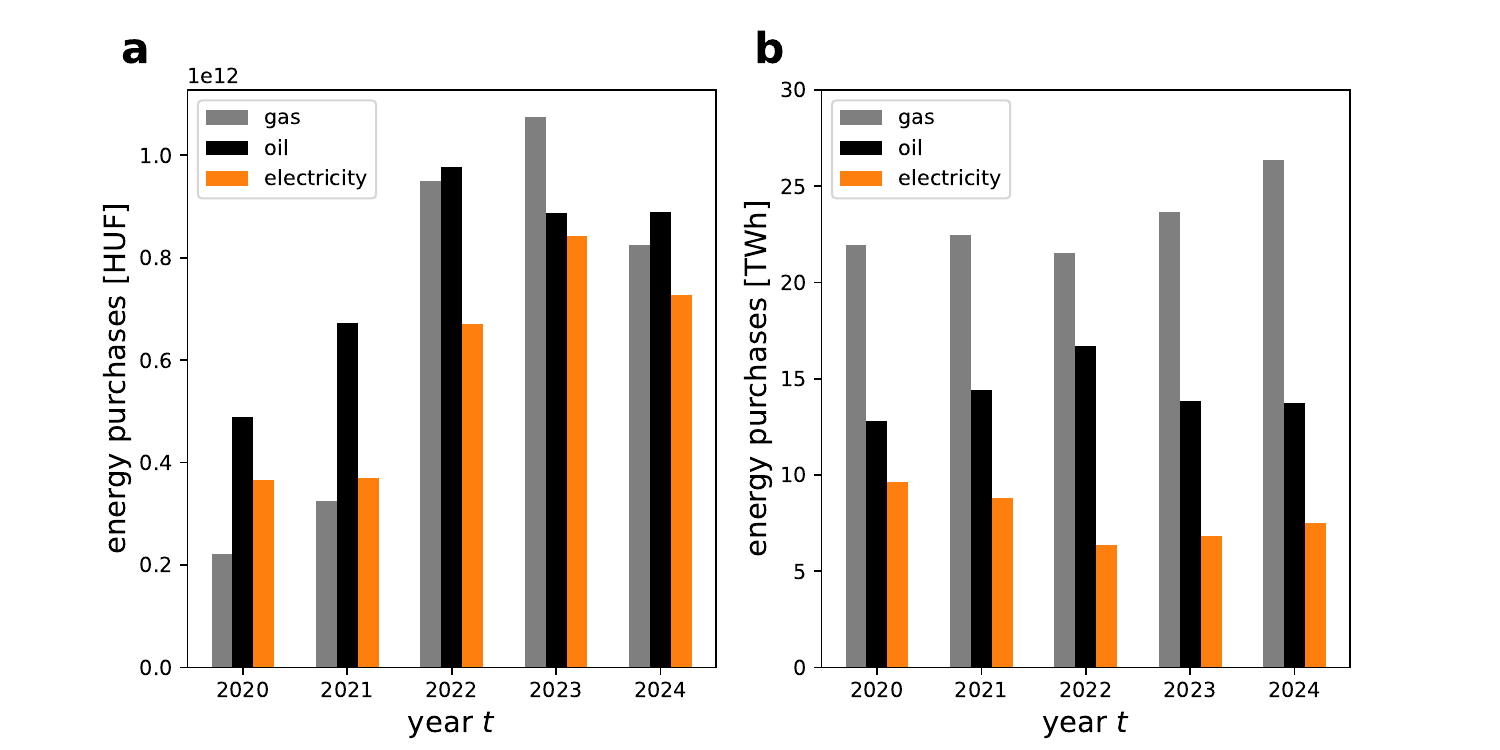}
\caption{Aggregated energy purchases in HUF and their conversion to terawatt-hours (TWh) for energy consumed by all firms in the sample, along with the calculated electricity shares for 2020-2024. (A) Aggregated monetary inputs for gas, electricity, and oil (in HUF) from 2020 to 2024. (B) Aggregated energy inputs for gas, electricity, and oil converted to terawatt-hours (TWh) for 2020-2024. (C) Calculated electricity share of total aggregated energy inputs in monetary terms. (D) Calculated electricity share of total aggregated energy inputs in terawatt-hours.}
\label{SI-FIG.:energy_purchases}
\end{figure*}

\begin{figure*}[htb]
\centering
\includegraphics[width=14.0cm]{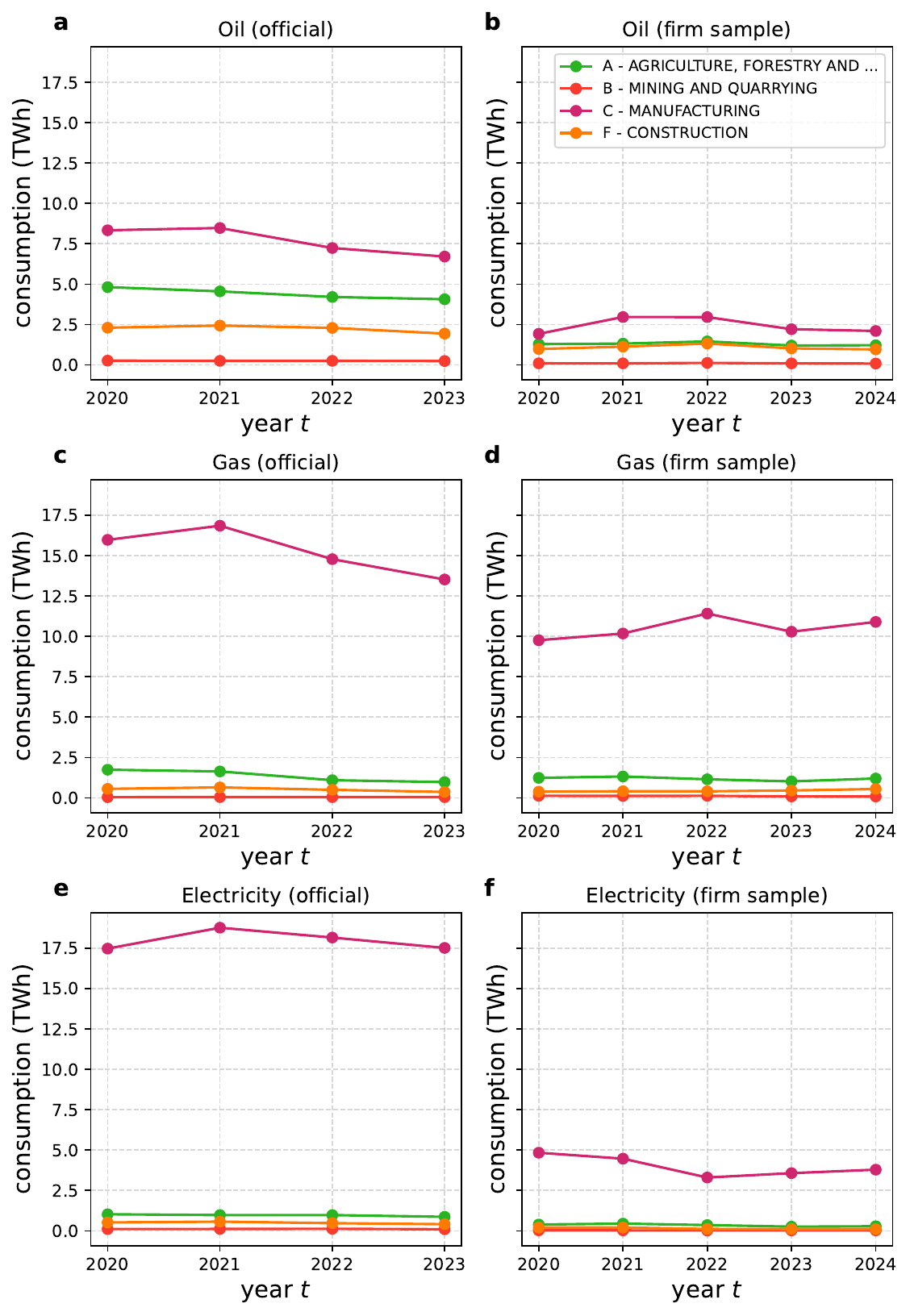}
\caption{Oil, gas, and electricity consumption by NACE 1-digit sectors ‘A – Agriculture, forestry and fishing,’ ‘B – Mining and quarrying,’ ‘C – Manufacturing,’ and ‘F – Construction,’ comparing Hungary’s official energy balances \cite{mekh_energy_balance_2025} with estimates from our aggregated firm sample. Panels show electricity consumption from official statistics (A) and the firm sample (B), gas consumption from official statistics (C) and the firm sample (D), and oil consumption from official statistics (E) and the firm sample (F).}
\label{SI-FIG.:energy_consumption_comparison}
\end{figure*}

\begin{figure*}[htb]
\centering
\includegraphics[width=16.0cm]{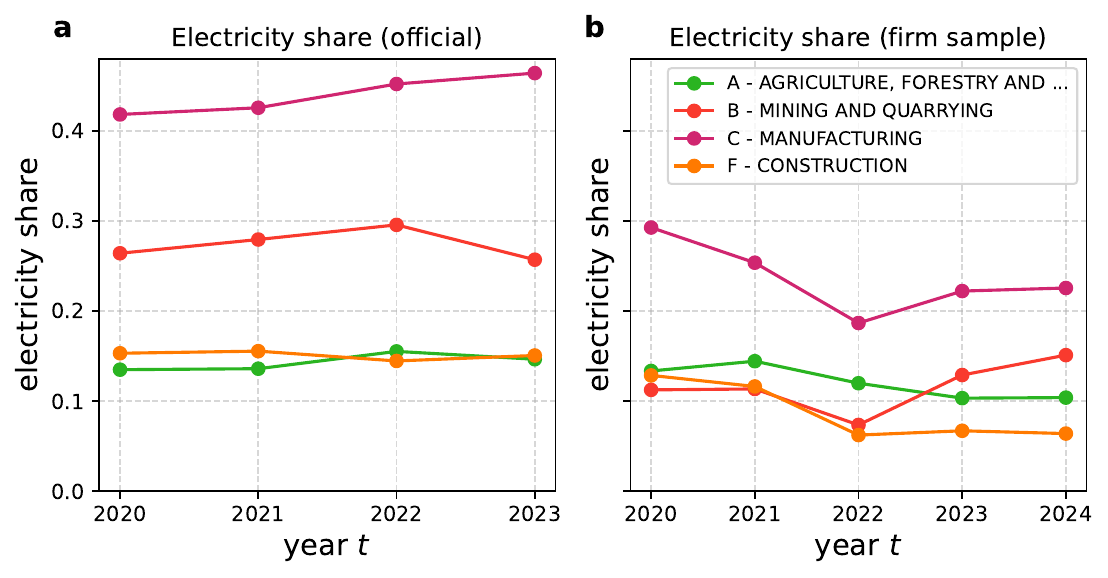}
\caption{Electricity shares of NACE 1-digit sectors, calculated from Hungary’s energy balance \cite{mekh_energy_balance_2025} for 2020-2023 compared with aggregated electricity shares of our firm sample. (A) Electricity shares for sectors 'A - Agriculture, forestry and fishing,' 'B - Mining and quarrying,' 'C - Manufacturing,' and 'F - Construction'. (B) Corresponding aggregated electricity shares of the same sectors in our firm sample.}
\label{SI-FIG.:electricity_shares_comparison}
\end{figure*}

\begin{figure*}[htb]
\centering
\includegraphics[width=16.0cm]{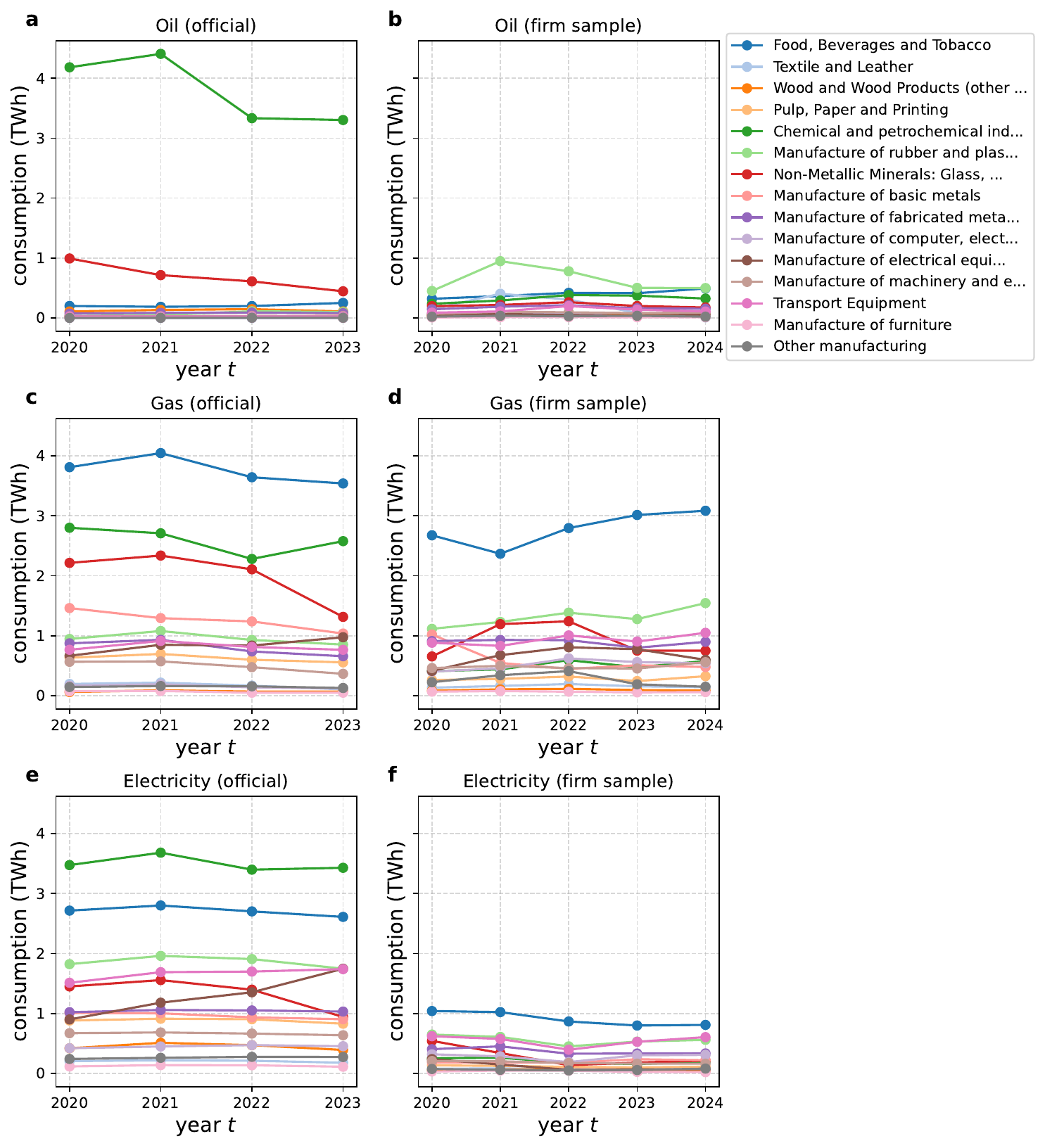}
\caption{Oil, gas, and electricity consumption by NACE 2-digit industries within the manufacturing sector 'C – Manufacturing', comparing Hungary’s official final energy use in industry for 2020-2023 \cite{mekh_industry_final_energy_2025} with estimates from our aggregated firm sample. Panels show electricity consumption from official statistics (A) and the firm sample (B), gas consumption from official statistics (C) and the firm sample (D), and oil consumption from official statistics (E) and the firm sample (F).}
\label{SI-FIG.:manufacuring_energy_consumption_comparison}
\end{figure*}

\begin{figure*}[htb]
\centering
\includegraphics[width=16.0cm]{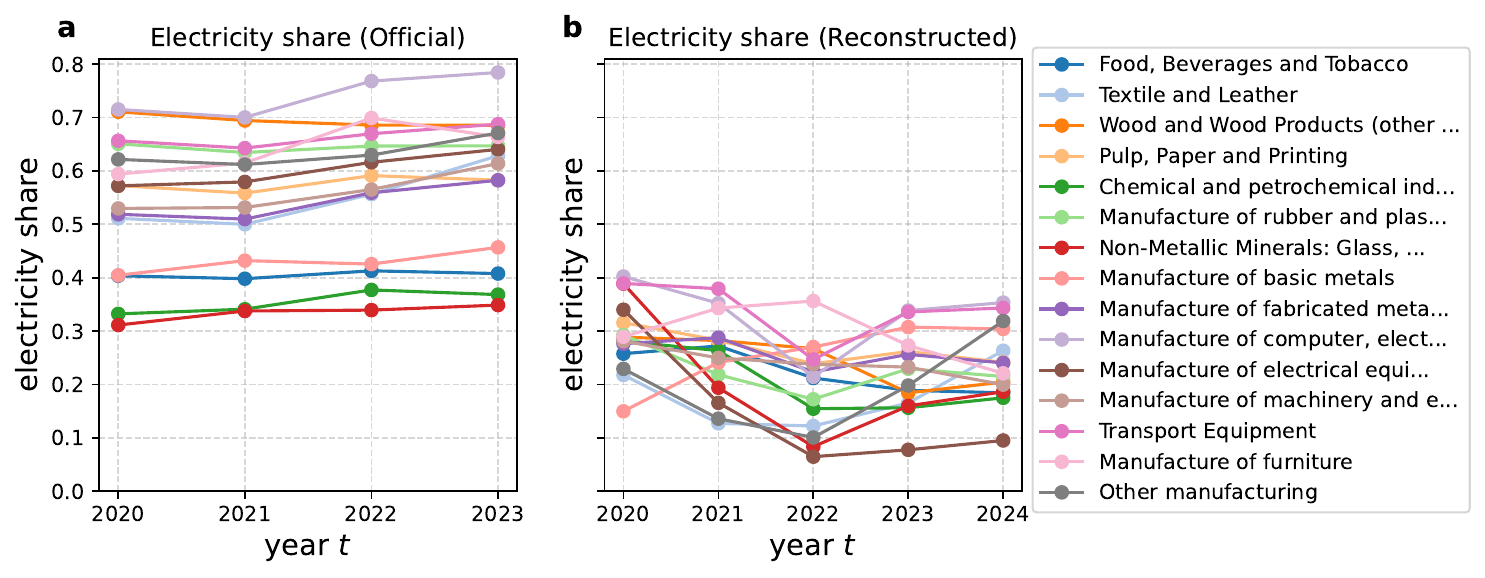}
\caption{Electricity shares by NACE 2-digit industries within the manufacturing sector 'C – ManufacturingÄ, comparing Hungary’s official final energy use in industry for 2020-2023 \cite{mekh_industry_final_energy_2025} with estimates from our aggregated firm sample. (A) Electricity shares for manufacturing sectors. (B) Corresponding aggregated electricity shares of the same sectors in our firm sample.}
\label{SI-FIG.:manufacuring_electricity_shares_comparison}
\end{figure*}

\begin{figure*}[htb]
\centering
\includegraphics[width=12.0cm]{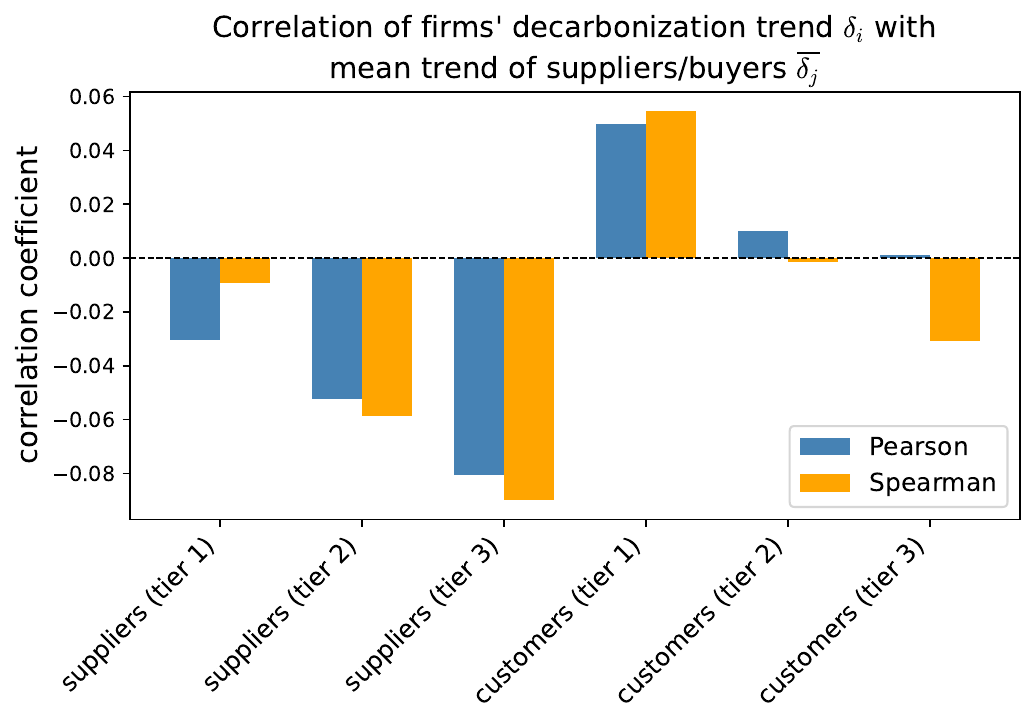}
\caption{Correlation between firms’ decarbonization trend and the mean trend of their suppliers and customers. For each tier (1-3), the decarbonization trend of a firm, $\delta_i$, is correlated with the mean trend of its suppliers and customers, $\overline{\delta_j}$. Pearson and Spearman correlation coefficients are reported separately for suppliers and customers across tiers.}
\label{SI-FIG.:transition_cascade_trend}
\end{figure*}

\begin{figure*}[htb]
\centering
\includegraphics[width=12.0cm]{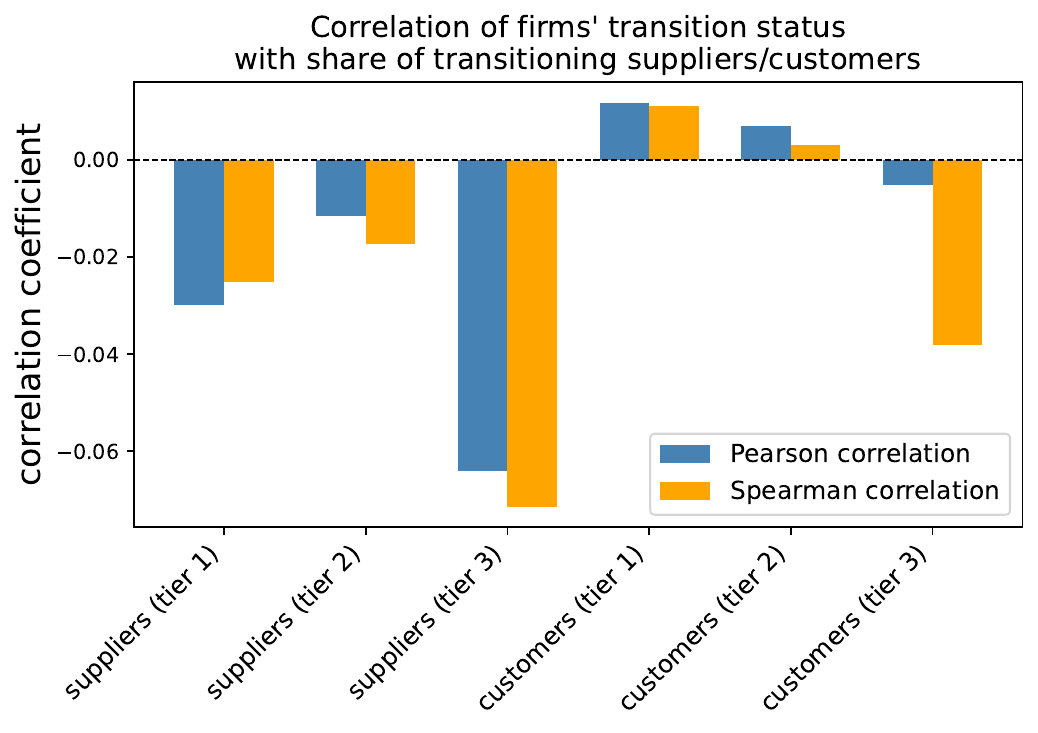}
\caption{Correlation between firms’ transition status (0/1) and the share of their transitioning suppliers and customers. Pearson and Spearman correlation coefficients are shown separately for suppliers and customers across tiers 1-3.}
\label{SI-FIG.:transition_cascade_share}
\end{figure*}

\begin{figure*}[htb]
\centering
\includegraphics[width=16.0cm]{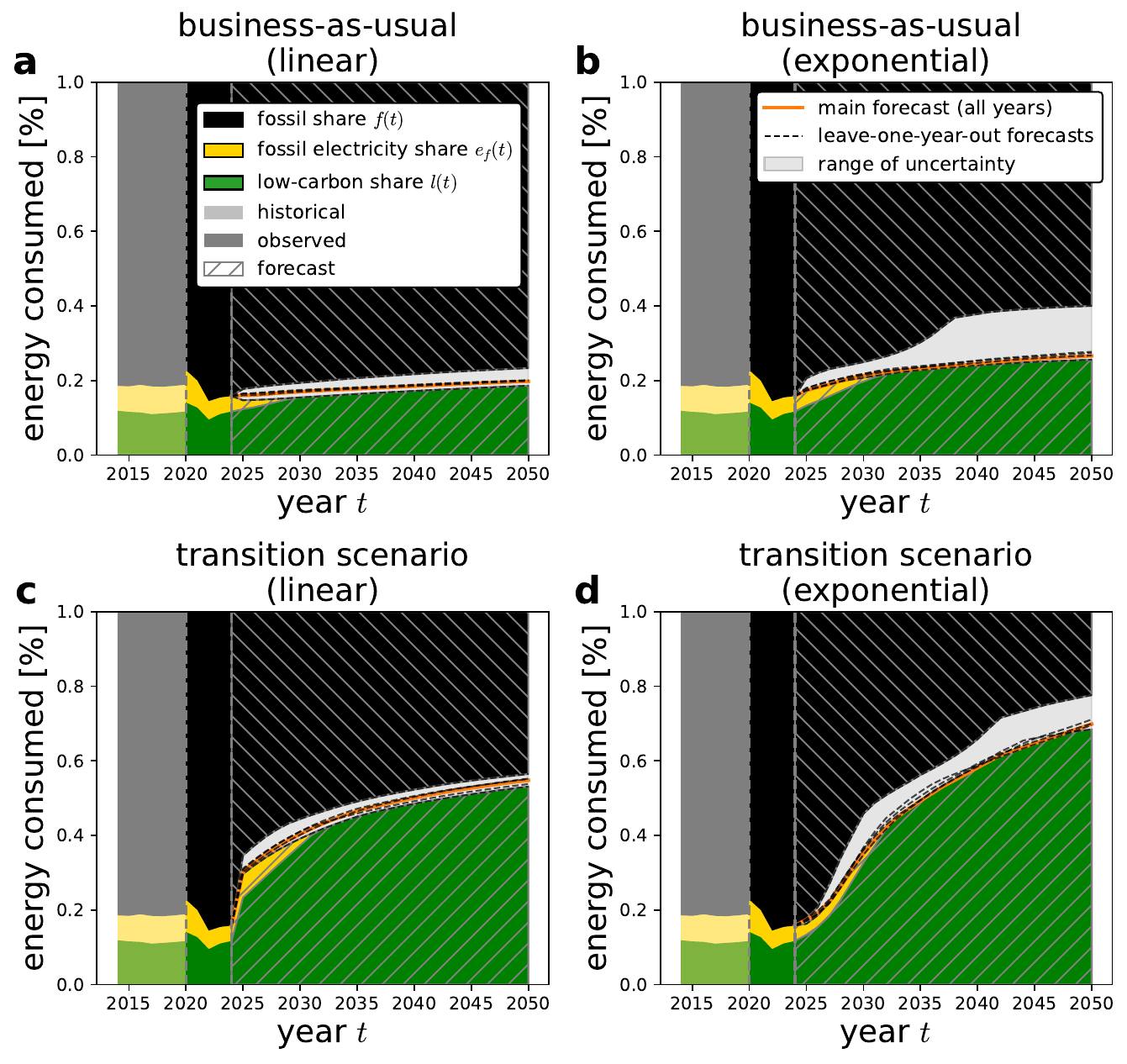}
\caption{Scenarios of aggregate energy consumption by fossil share, $f_(t)$, fossil electricity share, $e_f(t)$, and low-carbon share $l(t)$ based on observed firm-level trends. The gray shaded areas show the maximum uncertainty envelope across five runs per scenario: one main run using all years (2020--2024) and four additional runs where decarbonization trends or rates were re-estimated leaving out one year at a time. Panels show business-as-usual (linear, a; exponential, b) and transition scenarios (linear, c; exponential, d).}
\label{SI-FIG.:uncertainty_scenarios}
\end{figure*}

\FloatBarrier

\section*{Supplementary Tables}

\begin{table*}[htb]
\begin{tabular}{ccc}
\toprule
year & \makecell{low-carbon share \\ $u(t)$} & type \\
\midrule
2014 & 0.640014 & measured \\
2015 & 0.628091 & measured \\
2016 & 0.605395 & measured \\
2017 & 0.594775 & measured \\
2018 & 0.608424 & measured \\
2019 & 0.612310 & measured \\
2020 & 0.617984 & measured \\
2021 & 0.634349 & measured \\
2022 & 0.655953 & measured \\
2023 & 0.711712 & measured \\
2024 & 0.737821 & measured \\
2025 & 0.766675 & forecast \\
2026 & 0.798378 & forecast \\
2027 & 0.830082 & forecast \\
2028 & 0.861786 & forecast \\
2029 & 0.893489 & forecast \\
2030 & 0.925193 & forecast \\
2031 & 0.956897 & forecast \\
2032 & 0.988600 & forecast \\
2033 & 1.000000 & forecast \\
2034 & 1.000000 & forecast \\
2035 & 1.000000 & forecast \\
& & \\
2040 & 1.000000 & forecast \\
& & \\
2050 & 1.000000 & forecast \\
\bottomrule

\end{tabular}
\caption{Low-carbon share of the Hungarian electricity mix $u(t)$ for the years 2014-2050. Data for the years 2014-2024 are obtained from EMBER \cite{ember_electricity_2023}. The forecasts for the years 2024-2050 are based on a linear regression of the values for 2020-2024.}
\label{SI-Tab.:low-carbon_share_Hungary}
\end{table*}

\begin{table}
\label{tab:ratios}
\begin{tabular}{lr}
\toprule
variable [HUF] & \% covered \\
\midrule
gas input 2020 & 76.22 \\
gas input 2021 & 75.36 \\
gas input 2022 & 76.84 \\
gas input 2023 & 76.34 \\
gas input 2024 & 73.62 \\
electricity input 2020 & 71.74 \\
electricity input 2021 & 68.22 \\
electricity input 2022 & 66.46 \\
electricity input 2023 & 69.56 \\
electricity input 2024 & 70.66 \\
oil input 2020 & 52.66 \\
oil input 2021 & 50.38 \\
oil input 2022 & 51.50 \\
oil input 2023 & 49.45 \\
oil input 2024 & 48.91 \\
revenue 2020 & 42.97 \\
revenue 2021 & 42.84 \\
revenue 2022 & 42.70 \\
revenue 2023 & 42.70 \\
\bottomrule
\end{tabular}
\caption{Percentage coverage of the constructed firm sample in terms of aggregated gas, electricity, and oil inputs (in HUF), as well as total revenue, relative to the corresponding aggregated values for the total firm dataset from 2020 to 2024.}
\label{SI-Tab.:coverage_energy_inputs_revenue}
\end{table}

\begin{table*}[ht]
\centering
\begin{tabular}{l |cc|cc|cc|cc}
\hline
 & \multicolumn{2}{c|}{\textbf{\makecell{business-as-usual\\(linear)}}} & \multicolumn{2}{c|}{\textbf{\makecell{business-as-usual\\(exponential)}}} & \multicolumn{2}{c|}{\textbf{\makecell{transition scenario \\(linear)}}} & \multicolumn{2}{c}{\textbf{\makecell{transition scenario\\(exponential)}}} \\
\hline
{\rm year} & $l(t)$ & $f(t)$ & $l(t)$ & $f(t)$ & $l(t)$ & $f(t)$ & $l(t)$ & $f(t)$ \\
\hline
2020 & 0.141 & 0.772 & 0.141 & 0.772 & 0.141 & 0.772 & 0.141 & 0.772 \\
$2030$ & $0.170^{+0.024}_{-0.015}$ & $0.830^{+0.015}_{-0.024}$ & $0.213^{+0.036}_{-0.005}$ & $0.787^{+0.005}_{-0.036}$ & $0.403^{+0.041}_{-0.012}$ & $0.597^{+0.012}_{-0.041}$ & $0.353^{+0.105}_{-0.004}$ & $0.647^{+0.004}_{-0.105}$ \\
$2040$ & $0.185^{+0.032}_{-0.013}$ & $0.815^{+0.013}_{-0.032}$ & $0.245^{+0.133}_{-0.004}$ & $0.755^{+0.004}_{-0.133}$ & $0.500^{+0.022}_{-0.016}$ & $0.500^{+0.016}_{-0.022}$ & $0.582^{+0.076}_{-0.000}$ & $0.418^{+0.000}_{-0.076}$ \\
$2050$ & $0.198^{+0.035}_{-0.012}$ & $0.802^{+0.012}_{-0.035}$ & $0.266^{+0.134}_{-0.010}$ & $0.734^{+0.010}_{-0.134}$ & $0.548^{+0.017}_{-0.017}$ & $0.452^{+0.017}_{-0.017}$ & $0.699^{+0.078}_{-0.014}$ & $0.301^{+0.014}_{-0.078}$ \\

\end{tabular}
\caption{Aggregate low-carbon share, $l(t)$, and fossil share $f(t)$ shares across scenarios of Supplementary Fig. \ref{SI-FIG.:uncertainty_scenarios} with uncertainty bounds.}
\label{tab:scenarios_uncertainty}
\end{table*}

\begin{table*}
\begin{tabular}{ll}
  NACE 2-digit code & Sector labels in official statistic \\
  \hline
  10, 11, 12 & Food, Beverages and Tobacco \\
  13, 14, 15 & Textile and Leather \\
  16         & Wood and Wood Products (other than pulp and paper) \\
  17, 18     & Pulp, Paper and Printing \\
  19, 20, 21 & Chemical and petrochemical industries \\
  22         & Manufacture of rubber and plastic products \\
  23         & Non-Metallic Minerals: Glass, ceramic, cement and other building materials industries \\
  24         & Manufacture of basic metals \\
  25         & Manufacture of fabricated metal products, except machinery and equipment \\
  26         & Manufacture of computer, electronic and optical products \\
  27         & Manufacture of electrical equipment \\
  28         & Manufacture of machinery and equipment n.e.c. \\
  29, 30     & Transport Equipment \\
  31         & Manufacture of furniture \\
  32         & Other manufacturing \\
\end{tabular}
\caption{List of NACE 2-digit subsectors within ‘C – Manufacturing’. The sector labels from Hungary’s official final energy use in industry \cite{mekh_industry_final_energy_2025} were manually matched to NACE 2-digit codes. The correspondence is nearly one-to-one, but some sectors represent aggregates of multiple NACE 2-digit codes and are therefore presented here in aggregated form as well.}
\label{SI-Tab.:manufacturing_nace}
\end{table*}

\begin{table}[htbp]
\centering
\begin{tabular}{ll}
\hline
\textbf{Code} & \textbf{Description} \\
\hline
A & Agriculture, forestry and fishing \\
B & Mining and quarrying \\
C & Manufacturing \\
D & Electricity, gas, steam and air conditioning supply \\
E & Water supply; sewerage, waste management and remediation activities \\
F & Construction \\
G & Wholesale and retail trade; repair of motor vehicles and motorcycles \\
H & Transportation and storage \\
I & Accommodation and food service activities \\
J & Information and communication \\
L & Real estate activities \\
M & Professional, scientific and technical activities \\
N & Administrative and support service activities \\
O & Public administration and defence; compulsory social security \\
P & Education \\
Q & Human health and social work activities \\
R & Arts, entertainment and recreation \\
S & Other service activities \\
\hline
\end{tabular}
\caption{NACE 1-digit codes and their corresponding descriptions.}
\label{SI-Tab.:nace_codes}
\end{table}

\FloatBarrier

\begin{table}
\centering

\label{tab:self-consumption-1}

\begin{tabular}{ccccccccccccccc}
$\mathrm{year}\, t$ & \makecell{$C^{\mathrm{util}}(t)$\\(MW)} & \makecell{$C^{\mathrm{SCTE}}(t)$\\(MW)} & \makecell{$C^{\mathrm{HMKE}}(t)$\\(MW)} & \makecell{$C^{\mathrm{HMKE,res}}(t)$\\(MW)} & \makecell{$C^{\mathrm{HMKE,com}}(t)$\\(MW)} & \makecell{$G^{\mathrm{MEKH}}(t)$\\(GWh)} & \makecell{$G^{\mathrm{grid}}(t)$\\(GWh)} & \makecell{$G^{\mathrm{SC}}(t)$\\(GWh)} \\
\midrule
2020 & 1407.0 & 171.8 & 719.0 & 447.0 & 272.0 & 2459.0 & 1594.0 & 865.0 \\
2021 & 1829.3 & 225.9 & 1125.0 & 757.0 & 368.0 & 3796.0 & 2399.0 & 1397.0 \\
2022 & 2524.9 & 329.1 & 1492.0 & 1044.0 & 448.0 & 4732.0 & 3081.0 & 1651.0 \\
2023 & 3301.7 & 407.8 & 2329.0 & 1638.0 & 691.0 & 6925.0 & 4399.0 & 2526.0 \\
2024 & 4030.2 & 781.4 & 2540.0 & 1734.0 & 806.0 & 9200.0 & 5687.0 & 3513.0 \\
\\
source  & MAVIR \cite{MAVIR2025} & MAVIR \cite{MAVIR2025} & Forsense \cite{Forsense2024} & Forsense \cite{Forsense2024} & derived & MEKH-1 \cite{MEKH_1_2025} & \makecell{Energy Charts \cite{EnergyCharts2025} \\ ENTSO-E \cite{ENTSOE2025}} & derived \\
\bottomrule
\end{tabular}

\caption{Data and calculations to estimate commercial PV self-consumption for the years 2020-2024 [1]}

\end{table}

\begin{table}
\centering
\label{tab:self-consumption-2}
\begin{tabular}{ccccccccccccccc}
$\mathrm{year}\, t$ & $CF(t)$ & \makecell{$G^{\mathrm{SCTE}}(t)$\\(GWh)} & \makecell{$G^{\mathrm{HMKE,com}}(t)$\\(GWh)} & \makecell{$G^{\mathrm{SC,com}}(t)$\\(GWh)} & $s^{\mathrm{PV,com}}(t)$ & \makecell{$F^{\mathrm{total}}(t)$\\(GWh)} & \makecell{$F^{\mathrm{res}}(t)$\\(GWh)} & \makecell{$F^{\mathrm{com}}(t)$\\(GWh)} & $s^{\mathrm{com}}(t)$ \\
\midrule
2020 & 16.35\% & 246.7 & 389.6 & 636.3 & 25.88\% & 40046.0 & 12275.0 & 27771.0 & 2.29\% \\
2021 & 16.87\% & 333.8 & 543.8 & 877.7 & 23.12\% & 41716.0 & 12475.0 & 29241.0 & 3.00\% \\
2022 & 17.14\% & 672.7 & 672.7 & 1345.3 & 28.43\% & 41305.0 & 12348.0 & 28957.0 & 4.65\% \\
2023 & 16.66\% & 1008.5 & 1008.5 & 2016.9 & 29.13\% & 40197.0 & 12532.0 & 27665.0 & 7.29\% \\
2024 & 17.75\% & 1253.2 & 1253.2 & 2506.5 & 27.24\% & 40493.0 & 12089.0 & 28404.0 & 8.82\% \\
\bottomrule
\\
source  & MAVIR \cite{MAVIR2025} & derived & derived & derived & derived & MEKH-2 \cite{mekh_energy_balance_2025} & MEKH-2 \cite{mekh_energy_balance_2025} & derived & derived \\
\bottomrule
\end{tabular}

\caption{Data and calculations to estimate commercial PV self-consumption for the years 2020-2024 [2]}
\end{table}

\FloatBarrier

\end{document}